%% file: preprint-template.tex
\documentclass[twocolumn, switch]{article} 

\usepackage{preprint}

\usepackage{amsmath, amsthm, amssymb, amsfonts}

\usepackage{graphicx}
\usepackage{textcomp}
\usepackage{xcolor}
\usepackage[table]{xcolor}
\usepackage{listings}
\usepackage[numbers,square]{natbib}
\usepackage[utf8]{inputenc}	
\usepackage[T1]{fontenc}	
\usepackage[colorlinks = true,
            linkcolor = purple,
            urlcolor  = blue,
            citecolor = cyan,
            anchorcolor = black]{hyperref}	
\usepackage{booktabs} 		
\usepackage{nicefrac}		
\usepackage{microtype}		
\usepackage{lineno}		
\usepackage{float}			
\usepackage{comment}	
\usepackage{booktabs}
\usepackage{algorithm}
\usepackage{algpseudocode}
\usepackage{subcaption}
\usepackage[labelfont=bf,font=footnotesize]{caption}
\usepackage{url} 

\usepackage{newfloat}
\DeclareFloatingEnvironment[name={Supplementary Figure}]{suppfigure}
\usepackage{sidecap}
\sidecaptionvpos{figure}{c}

\usepackage{titlesec}
\titlespacing\section{0pt}{12pt plus 3pt minus 3pt}{1pt plus 1pt minus 1pt}
\titlespacing\subsection{0pt}{10pt plus 3pt minus 3pt}{1pt plus 1pt minus 1pt}
\titlespacing\subsubsection{0pt}{8pt plus 3pt minus 3pt}{1pt plus 1pt minus 1pt}

\usepackage{tikz,xcolor,hyperref}
\usetikzlibrary{shapes,arrows.meta,positioning,calc,matrix,decorations.pathreplacing, backgrounds}

\definecolor{lime}{HTML}{A6CE39}
\DeclareRobustCommand{\orcidicon}{
	\begin{tikzpicture}
	\draw[lime, fill=lime] (0,0)
	circle [radius=0.16]
	node[white] {{\fontfamily{qag}\selectfont \tiny ID}};
	\draw[white, fill=white] (-0.0625,0.095)
	circle [radius=0.007];
	\end{tikzpicture}
	\hspace{-2mm}
}
\foreach \x in {A, ..., Z}{\expandafter\xdef\csname orcid\x\endcsname{\noexpand\href{https://orcid.org/\csname orcidauthor\x\endcsname}
			{\noexpand\orcidicon}}
}

\title{FBench: A Flexible Benchmark for CFG-Based What-If Exploration of HPC I/O Patterns}

\usepackage{xwatermark}
\newwatermark[firstpage,color=gray!90,angle=0,scale=0.28, xpos=0in,ypos=-5in]{*correspondence: \texttt{zhu@uni-mainz.de}}

\usepackage{authblk}

\author[1\thanks{\tt{zhu@uni-mainz.de}}]{Zhaobin Zhu\orcidA{}}
\author[2]{Chen Wang\orcidB{}}
\author[3]{Kathryn Mohror\orcidC{}}
\author[1]{Sarah Neuwirth\orcidD{}}

\affil[1]{Johannes Gutenberg University Mainz, Germany}
\affil[2]{Nanyang Technological University, Singapore}
\affil[3]{Lawrence Livermore National Laboratory, USA}

\begin{document}
\definecolor{codegray}{rgb}{0.95,0.95,0.95}
\definecolor{keywordcolor}{rgb}{0.8,0,0.2}

\lstdefinestyle{mystyle}{
  backgroundcolor=\color{codegray},
  basicstyle=\ttfamily\footnotesize,
  keywordstyle=\color{keywordcolor}\bfseries,
  numbers=left,
  numberstyle=\tiny\color{gray},
  numbersep=6pt,
  frame=single,
  framerule=0.4pt,
  rulecolor=\color{black},
  showstringspaces=false,
  breaklines=true,
  keepspaces=true,
  captionpos=b,
  tabsize=2
}

\twocolumn[ 
  \begin{@twocolumnfalse} 

\maketitle

\begin{abstract}
The I/O performance of large-scale HPC applications depends on a complex interplay of access patterns, middleware optimizations, and file system configurations. To systematically explore these effects without repeatedly rerunning full applications, we introduce FBench, a flexible and code-transparent benchmarking tool for what-if analysis and I/O performance exploration. FBench leverages context-free grammars (CFGs) derived from Recorder traces to either generate simplified global configuration files for benchmark execution or replay I/O patterns on-the-fly without additional preprocessing. It supports both POSIX and MPI-IO interfaces and allows users to inject optimization hints via JSON configuration files, enabling rapid experimentation with I/O settings without code changes.
Our evaluation shows that FBench accurately reproduces I/O behavior for both synthetic and real workloads, capturing access patterns and performance trends across diverse optimizations and file system settings. For IOR and HACC-IO, FBench closely matches scaling behavior and sensitivity to Lustre striping parameters. For FLASH Sedov, it reveals that collective I/O on Lustre can yield up to $30\times$ lower write bandwidth than independent I/O, largely independent of striping, and that switching to a burst buffer file system increases non-collective write bandwidth by about $1.5\times$ without additional tuning. The evaluation with LAMMPS shows that FBench can significantly reduce the time required for what-if analyses and, with simple tuning, enable improvements of up to $8\times$. \looseness=-1

\end{abstract}
\vspace{0.35cm}

  \end{@twocolumnfalse} 
] 



\input{chapter/1_introduction}
\input{chapter/2_background}
\input{chapter/3_methodology}
\input{chapter/4_evaluation}
\input{chapter/5_conclusion}

\section*{Acknowledgments}
The authors gratefully acknowledge the computing time provided on the HPC systems at Lawrence Livermore National Laboratory. This work was performed under the auspices of the U.S. Department of Energy by Lawrence Livermore National Laboratory under Contract No. LLNL-CONF-2014268.
\normalsize
\bibliography{refs}


\end{document}

%% file: chapter/1_introduction.tex
\section{Introduction}
In the exascale era, HPC systems with hundreds of thousands to millions of processors and increasingly heterogeneous storage architectures pose significant challenges for application scalability. I/O performance is a major contributor, as I/O time can quickly become a limiting bottleneck \cite{carns2021understanding, lang2009performance, liu2012role, wang2014burstmem, behzad2014automatic}. Modern supercomputers rely on multi-tier storage architectures, ranging from flash-based burst buffers to disk and archival tiers, along with specialized data management libraries \cite{hdf5,netcdf,pnetCDF}, checkpointing mechanisms \cite{sato2014user, wang2015trio, wang2014burstmem}, and tailored file systems \cite{brim2023unifyfs, wang2022parallel, bent2009plfs}. As a result, application performance depends sensitively on three factors: the application’s access pattern, I/O library and middleware optimizations, and the configuration of the underlying file system \cite{bez2023access,bez2021bottleneck,bez2022drishti}. \looseness=-1


As I/O operations pass from high-level libraries through middleware to the parallel file system, the process quickly becomes complex. Interactions among these layers extend I/O behavior far beyond simple read and write operations, making it difficult to pinpoint where optimizations are most effective and how each layer can be tuned to overcome bottlenecks. Understanding I/O behavior, especially effects that emerge only at scale, requires detailed event traces \cite{mohror2009evaluating}.
For this purpose, applications are instrumented with tracing and profiling tools \cite{carns2021understanding, wang2020recorder, neuwirth2021parallel}, which capture more fine-grained I/O access behavior and expose potential inefficiencies. However, these tools focus less on performance measurement and do not provide a controlled environment for evaluating tuning parameters or exploring various I/O optimization strategies in practice.
Tools such as DXTExplorer \cite{bez2021bottleneck}, Drishti \cite{bez2022drishti}, and IOSIG \cite{yin2012boosting} can identify bottlenecks and suggest optimizations, but they offer no mechanism for evaluating these suggestions at the application level. Direct evaluation with real-world applications is both time- and resource-consuming, particularly when large codes must be repeatedly instrumented and executed for each configuration. \looseness=-1

What-if analysis is an essential approach to enable developers to study the performance impact of individual I/O design choices in isolation, without repeatedly modifying the source code and rerunning the full application. It allows alternative access patterns and tunable parameters to be explored quickly and systematically, and their impact to be assessed in a reproducible way. Unfortunately, existing tools lack a flexible and efficient way to translate profiling insights into measurable performance improvements without substantial overhead.
To avoid the cost of full application runs, benchmark-based what-if analyses are commonly used~\cite{dickson2016replicating}, but existing benchmarks are often either too specific or too generic and thus fail to capture realistic I/O patterns. Hence, time-consuming and error-prone application-specific I/O kernels often need to be derived manually~\cite{logan2011skel}. 
As a consequence, the gap between profiling and effective performance optimization remains largely unaddressed. On the one hand, it would improve the understanding of how I/O patterns influence performance. On the other hand, it would allow the impact of tunable parameters to be evaluated efficiently. \looseness=-1


In this work, 
we propose a flexible benchmark called \textit{FBench}, which can either interpret CFG-based traces on the fly or generate a configuration file from them to replay the I/O workload. Since both tunable parameters and fine-grained access patterns can be specified through the configuration file without requiring code changes to the original application, FBench provides an effective way to perform what-if analyses and efficiently explore the performance impact of different I/O patterns. Our key contributions are as follows: \looseness=-1

\begin{itemize}
    \item \textit{\textbf{CFG-based I/O Pattern Modeling}}: We use CFGs as an accurate and comprehensive representation of application I/O patterns. Our approach captures the essential characteristics of complex I/O access behavior in large-scale scientific applications. \looseness=-1
    
    \item \textbf{\textit{Accurate Performance Reproduction}}: We establish two ways to replicate an application’s I/O behavior. The first uses a simplified configuration file derived from the CFG to offer a global view. The second replays the full CFG on-the-fly, without pre-processing. Together, these approaches allow FBench to efficiently evaluate system I/O behavior while preserving performance fidelity comparable to the original application.
    
    \item \textit{\textbf{Application-agnostic What-if Analysis}}: We introduce a methodology for identifying and applying I/O optimization opportunities without domain knowledge or source code modifications, enabling pattern-level transformations rather than low-level code changes. By isolating I/O behavior, FBench enables fast what-if analysis and rapid exploration of optimizations.
    
\end{itemize}

%% file: chapter/2_background.tex
\section{Background and Related Work}


\subsection{CFG-based Pattern Representation}
A Context-Free Grammar is a formal notation for expressing recursive definitions of languages and is commonly used to define the syntax of programming languages. In its simplest form, a CFG can be viewed as a system of rule substitutions. Mathematically, CFG is defined as a 4-tuple~\cite{hopcroft2001introduction}: \[ G = (V, T, P, S), \]
which consists of the following components:
\begin{itemize}
    \item \textbf{Terminals ($T$)}: A finite set of symbols that appear in the actual strings of the language.
    \item \textbf{Variables / Non-Terminals ($V$)}: A finite set of symbols representing syntactic categories, with $V \cap T = \emptyset$.
    \item \textbf{Start Symbol ($S$)}: A distinguished variable $S \in V$ that represents the language being defined.
    \item \textbf{Productions ($P$)}: A finite set of rules. Each production has the form $A \rightarrow \alpha$, where $A \in V$ and $\alpha$ is a sequence of terminals and/or non-terminals.
\end{itemize}
As an example, consider the grammar $S \rightarrow 0 \mid 0S \mid 1S.$ In this case, the set of non-terminals is $V = \{S\}$, and the set of terminals is $T = \{0, 1\}$. The production rules are $P = \{\, S \rightarrow 0,\; S \rightarrow 0S,\; S \rightarrow 1S \,\}$ and the start symbol is $S$, which is also the only non-terminal. 
A key advantage of CFGs is their ability to represent traces as structured languages with nesting, repetition, and dependencies. Recurring patterns naturally become grammar rules, enabling structural compression that reduces storage while preserving semantic information. This makes CFGs well suited for capturing large-scale HPC execution traces, where repetitive patterns are common.
Modern tracing tools such as Siesta~\cite{luo2024siesta}, OmniscIO~\cite{dorier2014omnisc}, Recorder~\cite{wang2025recorder}, and Pilgrim~\cite{wang2021pilgrim} therefore employ CFG-based encodings to efficiently manage massive trace volumes. In this work, we use Recorder because it captures all function calls across targeted I/O stack layers and records complete parameter sets for each call. Such comprehensive logging produces large numbers of events, especially at scale. Recorder addresses this using CFGs combined with Call Signature Tables (CSTs), which map unique function signatures to terminal symbols. \looseness=-1


\begin{table}[h]
\centering
\caption{Example of a CFG rule and CST symbols.}
\begin{tabular}{|l|l|}
\hline
\textbf{CFG} & \textbf{CST} \\ \hline
$S \rightarrow a\ b$ & 
\begin{tabular}[c]{@{}l@{}}
a: \texttt{pwrite(fd, buf, 10, 0);}\\
b: \texttt{pwrite(fd, buf, 10, 10);}\\
\end{tabular} \\ \hline
\end{tabular}
\label{tab:cfg-cst}
\end{table}

As shown in Table \ref{tab:cfg-cst}, the CFG serves as a formal grammar with production rules that represent recurring calling patterns, while the CST functions as a hash table linking unique call signatures with terminal symbols. 
Recorder constructs CFGs online using the Sequitur algorithm, which is linear in the number of processed symbols \cite{wang2025recorder}. Because HPC I/O patterns often repeat frequently over time, CFG and CSTs based compression provides a compact, lossless representation of application I/O behavior suitable for both analysis and replay.

\subsection{Modeling and Emulating Application I/O Behavior}
Benchmarks and proxy applications are commonly used to evaluate I/O performance in HPC systems. However, developing these tools to accurately capture the application behavior requires substantial domain expertise, and many existing approaches struggle to balance fidelity, flexibility, and scalability. While synthetic benchmarks provide controlled parameter variation, proxy applications aim to emulate higher-level I/O semantics, serving as a bridge between low-level stress testing and full application execution.

A common approach among I/O benchmarks, such as FLASH-IO~\cite{FLASH-IO-Benchmark} and HACC-IO~\cite{hacc-io}, is to manually extract key I/O kernels from large-scale real-world applications. Both benchmarks focus primarily on checkpoint and restart operations, with FLASH-IO emphasizing write performance and HACC-IO extending this by supporting both read and write phases. This method effectively captures the I/O behavior of those specific applications, including their file sizes, checkpoint frequencies, and access patterns, but it lacks the flexibility to represent the broader variety of I/O behaviors and multi-phase workflows found across diverse scientific workloads. 
IOR~\cite{ior}, on the other hand, is a widely used synthetic, parameterized I/O benchmark. It addresses several limitations of I/O kernels, such as limited configurability and missing support for parallel libraries, by allowing users to vary access modes, transfer sizes, and interfaces. However, realistic application behavior is still difficult to reproduce with IOR, since real workloads typically consist of mixtures of different access patterns and file types, often interleaved over time, that cannot be captured by a single parameterized configuration.

Dickson et al.~\cite{dickson2016replicating} propose reproducing workloads through lightweight characterization by collecting I/O statistics with Darshan~\cite{Darshan} and configuring the MACSio~\cite{miller2015design} proxy accordingly. While this avoids rerunning the full application, characterization alone cannot capture complex access patterns, and MACSio’s reliance on high-level libraries limits its ability to represent diverse workloads accurately.
Behzad et al.~\cite{behzad2014automatic} use an early version of Recorder to in\-ter\-cept high-level I/O calls, aggregate per-process traces, and automatically generate compact SPMD code. This approach primarily targets HDF5 and does not generalize to other I/O interfaces. Moreover, modifying or exploring alternative access patterns requires regenerating code or changing the original application, limiting its usefulness for what-if analysis.
Skel~\cite{logan2011skel} generates skeletal I/O applications using ADIOS~\cite{adios_olcf}, with behavior defined via an external XML file. In contrast, our approach requires no manual configuration or external library: running the application is sufficient to obtain the CFG needed to reproduce its I/O workload.
Snyder et al.~\cite{snyder2015techniques} present an abstraction layer for trace-based, synthetic, and characterization-based workload modeling and compare their trade-offs. Their trace-based method, however, models only POSIX-level behavior and treats MPI largely as synchronization, preventing accurate reproduction of MPI-IO workloads. Additionally, the abstraction layer cannot generate workloads by itself and depends on external mechanisms for workload generation. 
Luo et al.~\cite{luo2017scalaioextrap} extend ScalaIOTrace~\cite{luo2015hpc} to extrapolate MPI I/O traces across scales and replay them in parallel while preserving per-rank semantics. However, the focus remains on reproducing original behavior, and any change to I/O patterns requires collecting new traces. Moreover, the availability of a standalone replay tool is unclear. \looseness=-1


Together, these efforts reveal a persistent gap: existing benchmarks are too rigid to express realistic multi-phase I/O behaviors, while proxy applications and trace-based replayers are limited by dependence on prior characterization, fixed access patterns, or external libraries. None of these approaches support flexible exploration of alternative I/O strategies without new tracing, reconfiguration, or manual intervention. This motivates the need for a more general, application-agnostic methodology capable of reproducing and varying I/O behavior with minimal developer effort, enabling systematic what-if analysis across storage configurations. \looseness=-1



%% file: chapter/3_methodology.tex
\section{The Flexible Benchmark -- FBench}

\subsection{Overview}
\begin{figure}[tb]
    \centering
    \resizebox{\columnwidth}{!}{
    \input{imgs/workflow_tk}
    }
    \caption{FBench workflow overview.}
    \label{fig:workflow}
    \vspace{-15pt}
\end{figure}
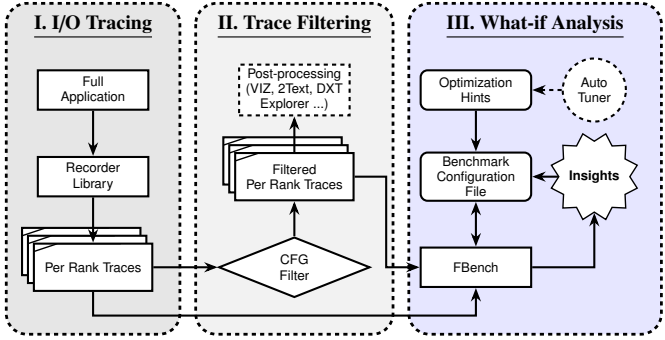

Figure~\ref{fig:workflow} illustrates the FBench workflow, consisting of three phases: (1) I/O tracing, (2) trace filtering, and (3) what-if analysis. Together, they provide a structured process for investigating, benchmarking, and exploring I/O performance optimizations in large-scale parallel applications.

\subsubsection{I/O Tracing Phase}
In the first phase, the target application is executed while instrumented or pre-loaded with the Recorder library. Unlike other characterization and tracing tools, Recorder intercepts all I/O calls across relevant layers and generates a per-rank trace in the form of a CFG, thereby preserving the structural relationships between operations. These traces capture detailed I/O behavior, including file paths, offsets, transfer sizes, timestamps, and call depth. With a runtime overhead of only ~3\%~\cite{wang2025recorder}, Recorder provides a favorable balance between trace detail and performance impact, making it practical for use in production-scale runs. Although comprehensive, raw traces often contain many low-impact operations, such as open calls without subsequent access, tiny metadata-related reads and writes, unused temporary files, extra \texttt{stat} or \texttt{seek} calls, or alignment-related transfers, that stem from internal library behavior rather than the intended application pattern, which should not be interpreted as part of the real application pattern.
 
\lstset{style=mystyle, language=C}

\begin{center}
\begin{minipage}{0.9\linewidth} 
\begin{lstlisting}[caption={Identical \texttt{pwrite()} calls resulting in distinct symbols in the CFG.}, label={lst:cfg-cst-pwrite}]
for (int i = 0; i < m; i++) {
    pwrite(fd, buf, size, offset);
}
\end{lstlisting}
\end{minipage}
\vspace{-10pt}
\end{center}

\subsubsection{Filtering Phase}
The second phase optionally filters or groups per-rank I/O operations. Filtering allows users to retain only operations relevant to performance analysis and discard incidental events. Since Recorder encodes each function call, including its arguments, as a unique CFG symbol, even semantically similar operations with different offsets or paths are represented as distinct symbols, as shown in Listing \ref{lst:cfg-cst-pwrite} and Table \ref{tab:cfg-cst}. Grouping such symbols or removing irrelevant ones reduces the CFG size and highlights the essential I/O behavior. The resulting filtered CFG remains valid inputs to post-processing tools such as Recorder-Viz, Drishti, and DXT-Explorer, thus often reduces post-processing time by decreasing the amount of data to interpret.

\begin{figure*}[t]
    \centering
    \resizebox{\textwidth}{!}{
    \input{imgs/merge_tk}
    }
    \caption{Transformation of per-rank traces into a structured FBench configuration via a global view. In this example, two files (\texttt{f1} and \texttt{f2}) are accessed as shared files, generating two configuration entries that involve all ranks. The first entry describes write operations to \texttt{f1} and contains two chunks corresponding to different transfer sizes: \texttt{s1} and \texttt{s2}. For each transfer size, the configuration aggregates how often the operation is executed across all ranks, resulting in a global view of \texttt{write$_{r_0 \dots r_{n-1}}$(f1, s1) $\times$ N} and \texttt{write$_{r_0 \dots r_{n-1}}$(f1, s2) $\times$ M}. Since \texttt{f1} is accessed only once, the total repeat count \texttt{k} is set to 1 accordingly. Similar to the first entry, the second entry represents read operations from \texttt{f2}, following the same structure but containing only a single chunk.}
    \label{fig:merging_per_rank_records}
    \vspace{-10pt}
\end{figure*}
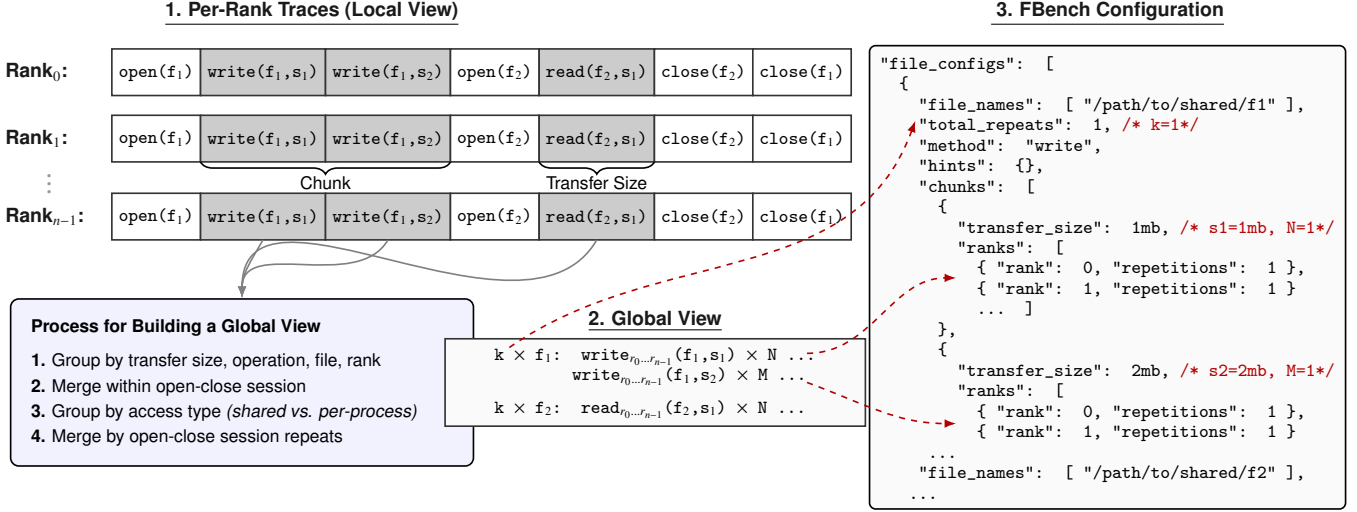

\subsubsection{What-If Analysis Phase}
The third phase enables exploration of alternative I/O configurations or system behaviors without rerunning the original application. FBench supports two execution modes: replaying I/O directly from the per-rank traces (raw or filtered) or using a benchmark configuration file, which can be written manually or automatically generated from the traces. In both cases, FBench interprets the trace or configuration and executes controlled benchmark runs, producing detailed performance metrics such as per-file bandwidth and aggregate throughput.


By default, FBench replays the actual I/O operations (call depth 1) when an application directly uses the native MPI-IO or POSIX interface. Since high-level I/O libraries ultimately translate their requests into MPI-IO or POSIX operations, their I/O behavior can also be reproduced by replaying the corresponding lower-level calls.
Based on benchmark results, optimization hints, such as buffer sizes, MPI-IO aggregation settings, or POSIX-level parameters, can be inserted into the configuration file. This creates an iterative feedback loop in which updated configurations are benchmarked until the desired performance characteristics are achieved. Because this workflow requires no modification of application code, FBench complements external what-if or autotuning tools such as Drishti and IOSIG: the optimization suggestions from these tools can be evaluated directly using FBench without rerunning the full application. \looseness=-1

FBench does not tune applications automatically. Instead, it provides an efficient environment for exploring potential optimizations and obtaining accurate performance estimates without executing the complete, time-consuming application workflow. To preserve predictive accuracy, FBench can insert synthetic delays during on-the-fly replay to approximate the temporal behavior of the original workload. In addition, the generated configuration file offers a transparent and structured view of the underlying I/O pattern, helping users understand how the application issues requests, identify inefficiencies in the access sequence, and reason about potential performance bottlenecks. \looseness=-1


\subsection{Simplified I/O Pattern Reconstruction (Global View)}

As shown in Figure \ref{fig:merging_per_rank_records}, FBench generates the configuration file from per-rank traces via a simplified global view. The construction of the global view follows steps analogous to CFG reconstruction, in which symbols and rules are merged. It consists of four main steps. First, all I/O operations are grouped by transfer size, operation type, file, and rank. Each unique combination becomes a configuration symbol. Second, configurations belonging to the same open-close session are merged, effectively combining repeated behavior within a session. Third, the resulting entries are grouped by access type, distinguishing shared-file from file-per-process I/O across different ranks and forming the chunk list for each configuration entry. Finally, configurations are merged based on the number of open-close sessions to determine the total number of repetitions for each configuration, yielding a compact, structured representation of the application’s I/O behavior.  \looseness=-1

By reducing a workload to summary quantities such as transferred volume and sustained bandwidth, the global view discards all temporal structure, including operation ordering, interleaving, and timing. This abstraction suffices for aggregate per-file throughput questions, as in steady checkpoint or dump phases and ior style microbenchmarking, where ranks behave near-identically and I/O phases are effectively temporally independent. Once temporal behavior itself becomes the object of analysis, this assumption no longer holds, and operation ordering, inter-operation gaps, burst structure, and collective load imbalance can no longer be recovered from global-view traces by construction. The temporal analysis presented above is exactly such a case. Consequently, for workloads with tight inter-process dependencies or highly asymmetric access patterns, the on-the-fly replay mode (Section~\ref{sec::on-the-fly}) preserves full per-rank fidelity where the global view cannot.

Based on these assumptions, the total data volume $D$ for a given file operation is computed as the sum over all chunks, where each chunk aggregates contributions from all participating ranks, multiplied by the total number of repetitions: \looseness=-1

\[
D = R_{\text{total}} \cdot \sum_{c \in \text{chunks}} \left( \sum_{r \in \text{ranks}(c)} \left( n_{r,c} \cdot s_c \right) \right)
\]
where $R_{\text{total}}$ is the number of open-close repetitions, $n_{r,c}$ is the number of repetitions performed by rank $r$ for chunk $c$, and $s_c$ is the transfer size of chunk $c$. 

From a complexity perspective, constructing the global configuration requires linear preprocessing. Let $R$ denote the number of ranks, $N$ the number of I/O records per rank, and $S$ the number of distinct open–close sessions. Processing and classifying the per-rank traces costs $\mathcal{O}(R \cdot N)$, and grouping sessions across ranks (for shared files, file-per-process I/O, and iteration-wise aggregation) adds $\mathcal{O}(S)$. Thus, the total preprocessing cost is $\mathcal{O}(R \cdot N + S)$, while executing the aggregated workload is proportional to the number of sessions, i.e., $\mathcal{O}(S)$. \looseness=-1

To keep FBench generic and flexible, and to enable the injection of hints and modification of access patterns for existing applications, FBench uses a JSON-formatted configuration for both benchmark execution and what-if analysis. FBench can automatically generate such configuration files directly from the global view representation, allowing users to reproduce or explore application I/O behavior without modifying the original code. A separate \texttt{file\_config} entry is created for each accessed file and operation type.  Each configuration specifies which file is accessed, how often the operation is performed, and which ranks participate. The optional \texttt{hints} field specifies optimization parameters, which are described in detail in Section~\ref{sec:passing_hints}. The \texttt{file\_names} entry lists the output file paths. For shared-file I/O this list contains a single path, while for file-per-process I/O it includes one path per process. The \texttt{total\_repeats} field defines how many open–close sessions are executed. Each session represents a complete I/O cycle in which the file is opened, the specified operations are performed, and the file is then closed again. The \texttt{method} field specifies the type of I/O operation. The \texttt{chunks} array defines the access pattern: each chunk represents a data block with a given transfer size, and the \texttt{ranks} list indicates which MPI ranks perform the writes and how many repetitions they execute. As a result, the transformation from per-rank traces to the global view enables a compact, repetition-based representation of application I/O behavior. \looseness=-1

\begin{algorithm}[tbp]
\caption{Time and Bandwidth Calculation}
\small
\begin{algorithmic}[1]
\For{each iteration, configuration}
  \State $t_s \gets \infty$, $t_e \gets 0$
  \For{each repeat}
    \If{rank participates}
      \State open file
      \If{first repeat} $t_s \gets \min(t_s,\ \texttt{MPI\_Wtime()})$ \EndIf
    \EndIf
    \State perform I/O
    \If{rank participates}
      \If{last repeat} $t_e \gets \max(t_e,\ \texttt{MPI\_Wtime()})$ \EndIf
      \State close file
    \EndIf
  \EndFor
  \State $bw = \text{bytes}/(\max_r t_e - \min_r t_s)$
\EndFor

\end{algorithmic}
\label{lst:calculate_time}
\end{algorithm}


Given the global view, the execution time can be determined by Algorithm~\ref{lst:calculate_time}, which computes the effective runtime of each configuration based on the earliest file-open time and the latest file-close time across all participating ranks. Each rank locally tracks its first timestamp~$t_s$ and last timestamp~$t_e$ over all repeats. After a configuration completes, a global reduction computes the earliest start time $T_s$ and the latest end time $T_e$ across ranks. The bandwidth is then computed by dividing the total amount of transferred data by the global I/O duration $T_e - T_s$. This measurement approach captures rank skew, variations in open and close times, and synchronization delays, and avoids the underestimation that occurs with per-rank timing, providing an accurate end-to-end throughput for the entire benchmark run. \looseness=-1

\subsection{On-the-Fly Replay (Local View)}
\label{sec::on-the-fly}
Beyond the configuration-based approach, FBench also provides an on-the-fly mode that replays I/O behavior directly from the per-rank traces. As each local CFG is read by the corresponding participating rank, every recorded event is immediately translated into a corresponding I/O operation, using the original arguments such as size, offset, and flags. This avoids decompressing or aggregating traces into a configuration file and enables a fine-grained reconstruction of the application’s behavior.
Each rank processes only its own trace, executing operations in the recorded order. With $R$ ranks and $N$ records per rank, the replay cost is $\mathcal{O}(R \cdot N)$, eliminating the preprocessing overhead of $\mathcal{O}(R \cdot N + S)$ required by the global configuration approach. \looseness=-1

\begin{figure}[tb]
    \centering
    \resizebox{\columnwidth}{!}{
    \input{imgs/per_rank_tk}
    }
    \caption{On-the-fly replay global synchronization using MPI-IO collective ops.}
    \label{fig:sync}
    \vspace{-10pt}
\end{figure}
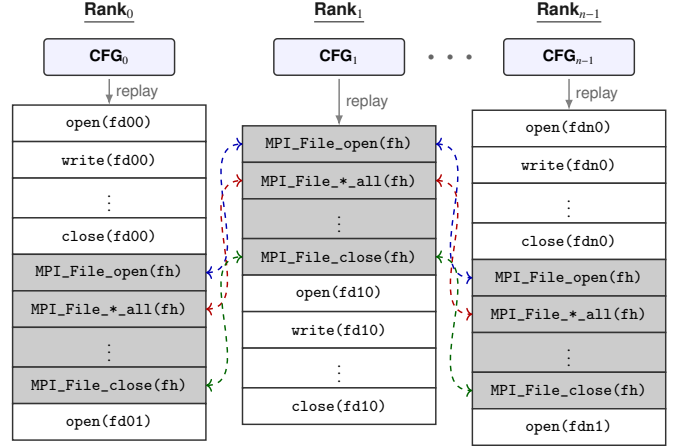


For pure POSIX I/O, there is no global temporal ordering between ranks, so replay proceeds independently on each process. In contrast, for MPI-IO the on-the-fly mode preserves synchronization semantics through collective operations such as \textit{MPI\_File\_open}, \textit{MPI\_File\_close}, and \textit{MPI\_File\_read/write\_*\_all}, as shown in Figure \ref{fig:sync}. Recorder encodes MPI file handles as globally unique integers, enabling FBench to match collective calls across all participating ranks and execute them consistently and in a coordinated manner. \looseness=-1

Compared to the simplified global view, the on-the-fly approach also replays additional file-access operations, including seeking, pointer updates, and offset-based reads and writes. Incorporating these operations allows FBench to reconstruct each rank’s exact access pattern and maintain correct file state throughout execution, which significantly improves the accuracy of fine-grained performance reproduction. Since the set of participating ranks for a collective operation can be inferred directly from the globally encoded handles, no explicit creation of sub-communicators is required, further reducing implementation and setup overhead. \looseness=-1

Temporal behavior can also be preserved when needed. For analyses where timing is important (e.g., frequency- or runtime-based studies), omitting delays between I/O events would distort the original timing structure, leading to unrealistic load bursts and skewed latency measurements. During on-the-fly replay, FBench can insert inter-event delay. By default, the time difference between the end of event $e_1$ and the start of event $e_2$ is computed from the trace and applied immediately before executing $e_2$. This maintains a realistic execution timeline while still enabling controlled benchmarking.\looseness=-1


\subsection{Passing Optimization Hints}
\label{sec:passing_hints}
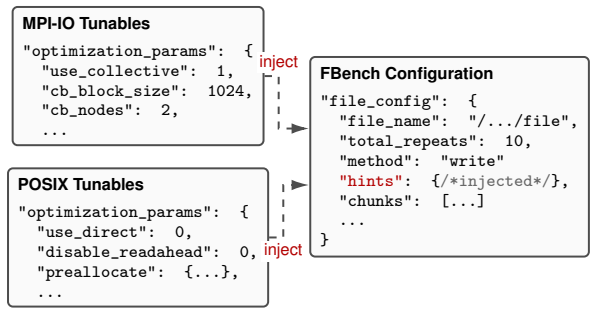
\begin{figure}[tb]
    \centering
    \resizebox{0.9\columnwidth}{!}{
    \input{imgs/hints_tk}
    }
    \caption{Injection of MPI-IO and POSIX optimization parameters into the FBench configuration, enabling backend-specific tuning without code changes.}
    \label{fig:passing_optimization}
\end{figure}

To enable what-if analysis without changing application code, FBench allows optimization parameters and I/O patterns to be modified at replay time. As shown in Figure~\ref{fig:passing_optimization}, FBench uses a JSON configuration file as input, access patterns and hints can be edited easily and in a human-readable way. This lets researchers and operators experiment with alternative storage and MPI-IO setups by changing a single configuration artifact rather than rebuilding or instrumenting the full application. \looseness=-1

Optimization parameters can be specified for both POSIX I/O and MPI-IO. These hints allow the replay engine to control file-system behavior, buffering strategies, and access optimizations without modifying the underlying system, enabling different performance scenarios to be evaluated in a controlled and reproducible way. Due to the separation of the replay engine and the injection of tunables, FBench can be easily extended with further optimization parameters.

On the POSIX level, hints influence how standard file operations are handled at the system level. Settings such as \texttt{use\_direct} and \texttt{disable\_readahead} control page-cache and read-ahead behavior, exposing the application’s raw access pattern. \texttt{preallocate} reserves file space up front to reduce fragmentation, while file-system-specific options, such as disabling \texttt{atime} updates, relaxing lock handling, and configuring stripe size and stripe count, directly affect metadata overhead and parallelism.

MPI-IO parameters control ROMIO’s internal optimizations. Specifying the expected access pattern and enabling or disabling collective I/O determines whether ROMIO issues independent operations or applies its two-phase I/O algorithm. Parameters such as \texttt{cb\_block\_size}, \texttt{cb\_buffer\_size}, and \texttt{cb\_nodes} shape collective buffering, while options like \texttt{romio\_cb\_read/write} and data-sieving settings determine how ROMIO merges and reorganizes individual requests. Buffer sizes for independent reads and writes define how much temporary memory ROMIO may use for noncontiguous access patterns. In addition, file-system-specific options analogous to the POSIX hints can be passed through the MPI-IO configuration, allowing different striping configurations and I/O strategies to be evaluated directly during MPI-IO replay.

%% file: imgs/workflow_tk.tex
\begin{tikzpicture}

\tikzset{
    process/.style={
        rectangle, draw=black, line width=1.5pt, fill=white,
        minimum width=2.8cm, minimum height=1cm, align=center, font=\sffamily, inner sep=3pt
    },
    note/.style={
        rectangle, draw=black, line width=1.5pt, fill=white,
        minimum width=2.8cm, minimum height=1cm, align=center, font=\sffamily, rounded corners=5pt, inner sep=3pt
    },
    decision/.style={
        diamond, draw=black, line width=1.5pt, fill=white, aspect=2.5,
        minimum width=2.2cm, minimum height=1.3cm, text width=2cm, align=center, font=\sffamily, inner sep=1pt
    },
    starburst/.style={
        star, star points=12, star point ratio=0.8, draw=black, line width=1pt, fill=white,
        minimum size=1.8cm, align=center, font=\sffamily\bfseries, inner sep=0pt
    },
    dashedcircle/.style={
        circle, draw=black, line width=1.5pt, dashed, fill=white,
        minimum size=1.6cm, align=center, font=\sffamily, inner sep=1pt
    },
    dashedprocess/.style={
        rectangle, draw=black, line width=1.5pt, dashed, fill=white,
        minimum width=2.8cm, minimum height=1.2cm, align=center, font=\sffamily, inner sep=3pt
    },
    folder/.style={
        rectangle, draw=black, line width=1.5pt, fill=white,
        minimum width=3.0cm, minimum height=1.2cm, align=center, inner sep=3pt, font=\sffamily,
        path picture={
            \draw[line width=1pt] ($(path picture bounding box.north west)+(0,-0.2)$) -- ($(path picture bounding box.north west)+(0.6,0)$) -- ($(path picture bounding box.north east)+(-0.8,0)$);
        }
    },
    container/.style={
        draw=black, line width=2pt, dashed, rounded corners=15pt
    },
    arrow/.style={ ->, >=Stealth, line width=1.5pt },
    dashedarrow/.style={ ->, >=Stealth, line width=1.5pt, dashed },
    biarrow/.style={ <->, >=Stealth, line width=1.5pt }
}

    \begin{pgfonlayer}{background}
        \draw[container, fill=black!10] (-2.2, 2.2) rectangle (2.2, -6.2);
        \draw[container, fill=black!5] (2.6, 2.2) rectangle (7.6, -6.2);
        \draw[container, fill=blue!10] (8.0, 2.2) rectangle (14.4, -6.2);
    \end{pgfonlayer}

    \node (h1) at (0, 1.6) {\Large\bfseries\underline{I. I/O Tracing}};
    \node [process] (app) at (0, 0) {Full\\Application};
    \node [process] (lib) at (0, -2.2) {Recorder\\Library};
    
    \node [folder] (perRank3) at (-0.3, -4.1) {};
    \node [folder] (perRank2) at (-0.15, -4.3) {};
    \node [folder] (perRank1) at (0, -4.5) {Per Rank Traces};

    \node (h2) at (5.1, 1.6) {\Large\bfseries\underline{II. Trace Filtering}};
    \node [dashedprocess] (post) at (5.1, 0) {Post-processing\\(VIZ, 2Text, DXT\\Explorer ...)};
    
    \node [folder] (filtered3) at (4.8, -1.8) {};
    \node [folder] (filtered2) at (4.95, -2.0) {};
    \node [folder] (filtered1) at (5.1, -2.2) {Filtered \\ Per Rank Traces};
    
    \node [decision] (cfg) at (5.1, -4.5) {CFG\\Filter};

    \node (h3) at (11.2, 1.6) {\Large\bfseries\underline{III. What-if Analysis}};
    
    \node [note] (hints) at (9.7, 0) {Optimization\\Hints}; 
    \node [note] (config) at (9.7, -2.2) {Benchmark\\Configuration\\File};
    \node [process] (fbench) at (9.7, -4.5) {FBench};

    \node [dashedcircle] (tuner) at (12.7, 0) {Auto\\Tuner};
    \node [starburst] (insights) at (12.7, -2.2) {Insights};

    
    \draw [arrow] (app) -- (lib);
    \draw [arrow] (lib) -- (perRank1.north);
    
    \draw [arrow] (filtered1.north) -- (post.south);
    \draw [arrow] (cfg.north) -- (filtered1.south);
    
    \draw [arrow] (hints) -- (config);
    \draw [biarrow] (config) -- (fbench);
    \draw [dashedarrow] (tuner) -- (hints);
    
    \draw [arrow] (insights.west) -- (config.east);
    
    \draw [arrow] (fbench.east) -| (insights.south);
    
    \draw [arrow] (perRank1.east) -- (cfg.west);

    \draw [arrow] (filtered1.east) -- ++(0.8,0) |- ($(fbench.west)$);
    \draw [arrow] (perRank1.south) -- ++(0,-0.6) -| ($(fbench.south)$);

\end{tikzpicture}

%% file: imgs/merge_tk.tex
\begin{tikzpicture}[
    font=\sffamily\small,
    >=Stealth,
    record/.style={
        draw=black!80, thick, fill=white,
        minimum width=1.2cm, minimum height=0.8cm, align=center,
        font=\ttfamily\small, outer sep=0pt
    },
    record short/.style={
        record, minimum width=.8cm
    },
    config/.style={
        draw=black, thick, fill=black!2, rounded corners=3pt,
        minimum width=5cm, minimum height=7cm, align=left,
        inner sep=5pt, font=\ttfamily\small,
    },
    title/.style={
        font=\sffamily\bfseries, text=black!90
    },
    pipeline/.style={
        draw=black, thick, fill=blue!5!,
        rounded corners=4pt, align=left,
        font=\sffamily\small, inner sep=10pt
    },
    arrow/.style={
        -{Latex[length=3mm, width=2mm]}, thick, draw=black!60
    },
    step label/.style={
        font=\sffamily\bfseries\small, color=black!70!black,
    },
    connect arrow/.style={
        -{Latex[length=2mm, width=1.5mm]}, thick, dashed
    },
    connect records/.style={
        -{Latex[length=2mm, width=1.5mm]}, thick
    },
    rank/.style={
        font=\sffamily\bfseries, text=black!90, align=left, text width=1.5cm
    }
]

\node[title] (left_title) at (4.5, 2.5) {\underline{1. Per-Rank Traces (Local View)}};

\node[rank] (r0_label) at (0, 1.5) {Rank$_0$:};
\node[record short, right=0.2cm of r0_label] (r0_1) {open(f$_1$)};
\node[record, fill=black!20, right=0pt of r0_1] (r0_2) {write(f$_1$,s$_1$)};
\node[record, fill=black!20, right=0pt of r0_2] (r0_3) {write(f$_1$,s$_2$)};
\node[record short, right=0pt of r0_3] (r0_4) {open(f$_2$)};
\node[record, fill=black!20, right=0pt of r0_4] (r0_5) {read(f$_2$,s$_1$)};
\node[record short, right=0pt of r0_5] (r0_6) {close(f$_2$)};
\node[record short, right=0pt of r0_6] (r0_7) {close(f$_1$)};

\node[rank, below=0.6cm of r0_label] (r1_label) {Rank$_1$:};
\node[record short, right=0.2cm of r1_label] (r1_1) {open(f$_1$)};
\node[record, fill=black!20, right=0pt of r1_1] (r1_2) {write(f$_1$,s$_1$)};
\node[record, fill=black!20, right=0pt of r1_2] (r1_3) {write(f$_1$,s$_2$)};
\node[record short, right=0pt of r1_3] (r1_4) {open(f$_2$)};
\node[record, fill=black!20, right=0pt of r1_4] (r1_5) {read(f$_2$,s$_1$)};
\node[record short, right=0pt of r1_5] (r1_6) {close(f$_2$)};
\node[record short, right=0pt of r1_6] (r1_7) {close(f$_1$)};

\node[font=\Large\bfseries, text=gray!80, below=0 cm of r1_label] {$\vdots$};

\node[rank, below=0.8cm of r1_label] (rn_label) {Rank$_{n-1}$:};
\node[record short, right=0.2cm of rn_label] (rn_1) {open(f$_1$)};
\node[record, fill=black!20, right=0pt of rn_1] (rn_2) {write(f$_1$,s$_1$)};
\node[record, fill=black!20, right=0pt of rn_2] (rn_3) {write(f$_1$,s$_2$)};
\node[record short, right=0pt of rn_3] (rn_4) {open(f$_2$)};
\node[record, fill=black!20, right=0pt of rn_4] (rn_5) {read(f$_2$,s$_1$)};
\node[record short, right=0pt of rn_5] (rn_6) {close(f$_2$)};
\node[record short, right=0pt of rn_6] (rn_7) {close(f$_1$)};

\draw[decorate, decoration={brace, amplitude=6pt, mirror}, thick]
    (r1_2.south west) -- (r1_3.south east) node[midway, below=3pt] {Chunk};

\draw[decorate, decoration={brace, amplitude=6pt, mirror}, thick]
    (r1_5.south west) -- (r1_5.south east) node[midway, below=3pt] {Transfer Size};

\node[pipeline, below=1cm of rn_3, anchor=north, xshift=-2.5cm] (pipeline) {
    \textbf{Process for Building a Global View}\\[1.5ex]
    \textbf{1.} Group by transfer size, operation, file, rank\\[0.5ex]
    \textbf{2.} Merge within open-close session \\[0.5ex]
    \textbf{3.} Group by access type \textit{(shared vs. per-process)}\hspace*{0.6cm} \\[0.5ex]
    \textbf{4.} Merge by open-close session repeats
};

\node[record, minimum width=7.2cm, minimum height=1.5cm, fill=black!2, draw=black!80, inner sep=3pt, align=left, anchor=west] 
(m1) at ($(pipeline.east) + (-0.5cm, 0)$) {
   \textbf{k $\times$ f$_1$}: \textbf{write$_{r_0 \dots r_{n-1}}$(f$_1$,s$_1$) $\times$ N} \dots  \\ ~~~~~~~ \textbf{write$_{r_0 \dots r_{n-1}}$(f$_1$,s$_2$) $\times$ M} \dots\\[1.5ex]
   \textbf{k $\times$ f$_2$}: \textbf{read$_{r_0 \dots r_{n-1}}$(f$_2$,s$_1$) $\times$ N} \dots
};

\node[title, above=0cm of m1.north]{\underline{2. Global View}};

\node[config, anchor=north west] (config_doc) at ($(rn_7.east |- left_title.south) + (0.3cm, -0.2cm)$) {

"file\_configs": [\\
~~\{ \\
~~~~"file\_names": [ "/path/to/shared/f1" ],\\
~~~~"total\_repeats": 1,         \textcolor{red!70!black}{/* k=1*/} \\
~~~~"method": "write",\\
~~~~"hints": \{\},\\
~~~~"chunks": [\\
~~~~~~\{\\
~~~~~~~~{"transfer\_size"}: 1mb, \textcolor{red!70!black}{/* s1=1mb, N=1*/}\\
~~~~~~~~"ranks": [\\
~~~~~~~~~~\{ "rank": 0, "repetitions": 1 \},\\
~~~~~~~~~~\{ "rank": 1, "repetitions": 1 \}\\
~~~~~~~~~~... ]\\
~~~~~~\},\\
~~~~~~\{\\
~~~~~~~~{"transfer\_size"}: 2mb, \textcolor{red!70!black}{/* s2=2mb, M=1*/}\\
~~~~~~~~"ranks": [\\
~~~~~~~~~~\{ "rank": 0, "repetitions": 1 \},\\
~~~~~~~~~~\{ "rank": 1, "repetitions": 1 \}\\
~~~~~... \\
~~~~"file\_names": [ "/path/to/shared/f2" ],\\
~~ ...
};

\node[title, above=0.2cm of config_doc] {\underline{ 3. FBench Configuration}};

\coordinate (stack_t0) at ($(m1.center) + (-2.5cm, 0.6cm)$);
\coordinate (stack_t1) at ($(m1.center) + (2.6cm, 0.5cm)$);
\coordinate (stack_t2) at ($(m1.center) + (2.6cm, -0cm)$);

\coordinate (json_t0) at ($(config_doc.north west) + (0.8cm, -1.3cm)$);
\coordinate (json_t1) at ($(config_doc.north west) + (1.5cm, -4cm)$);
\coordinate (json_t2) at ($(config_doc.north west) + (1.5cm, -6.5cm)$);

\draw[connect arrow, red!70!black] (stack_t0) to[out=35, in=250] (json_t0);
\draw[connect arrow, red!70!black] (stack_t1) to[out=0, in=180] (json_t1);
\draw[connect arrow, red!70!black] (stack_t2) to[out=340, in=180] (json_t2);

\draw[connect records, black!50] (rn_2.south) to[out=240, in=90] (pipeline.north);
\draw[connect records, black!50] (rn_3.south) to[out=240, in=90] (pipeline.north);
\draw[connect records, black!50] (rn_5.south) to[out=240, in=90] (pipeline.north);

\end{tikzpicture}

%% file: imgs/per_rank_tk.tex
\begin{tikzpicture}[
        posix/.style={
            draw=black!80, 
            thick, 
            minimum width=3.7cm, 
            minimum height=0.7cm, 
            align=center, 
            fill=white, 
            font=\ttfamily\small,
            outer sep=0pt
        },
        mpi/.style={
            draw=black!80, 
            thick, 
            minimum width=3.7cm, 
            minimum height=0.7cm, 
            align=center, 
            fill=black!20, 
            font=\ttfamily\small\bfseries, 
            outer sep=0pt
        },
        sync/.style={
            <->, 
            dashed, 
            thick, 
            draw=blue!70!black 
        },
        sync2/.style={
            <->, 
            dashed, 
            thick, 
            draw=red!70!black 
        },
        sync3/.style={
            <->, 
            dashed, 
            thick, 
            draw=green!40!black 
        },
        header/.style={
            font=\sffamily\bfseries,
            text=black!90
        },
        cfg/.style={
            draw=black!80, 
            thick, 
            rounded corners=2pt, 
            fill=blue!5!,    
            minimum width=2.5cm, 
            minimum height=0.7cm, 
            align=center, 
            font=\sffamily\small\bfseries,
            outer sep=2pt
        }
    ]
    
    \coordinate (C0) at (0,0);
    \coordinate (C1) at (4.4,0); 
    \coordinate (C2) at (8.8,0); 

    \node[header] (h0) at (C0) {\underline{Rank$_0$}};
    
    \node[cfg, below=0.1cm of h0] (cfg0) {CFG$_0$};
    
    \node[posix, below=0.5cm of cfg0] (r0_1) {open(fd00)};
    
    \draw[-{Latex[length=2mm, width=1.5mm]}, thick, draw=black!50] (cfg0) -- (r0_1) node[midway, right, font=\small\sffamily, text=black!60] {replay};
    
    \node[posix, below=0pt of r0_1] (r0_2) {write(fd00)};
    \node[posix, below=0pt of r0_2] (r0_3) {$\vdots$};
    \node[posix, below=0pt of r0_3] (r0_4) {close(fd00)};
    
    \node[mpi, below=0pt of r0_4]   (r0_5) {MPI\_File\_open(fh)};
    \node[mpi, below=0pt of r0_5]   (r0_6) {MPI\_File\_*\_all(fh)};
    \node[mpi, below=0pt of r0_6]   (r0_7) {$\vdots$};
    \node[mpi, below=0pt of r0_7]   (r0_8) {MPI\_File\_close(fh)};
    
    \node[posix, below=0pt of r0_8] (r0_9) {open(fd01)};
    
    \node[header] (h1) at (C1) {\underline{Rank$_1$}};
    
    \node[cfg, below=0.1cm of h1] (cfg1) {CFG$_1$};
    
    \node[mpi, below=0.9cm of cfg1] (r1_1) {MPI\_File\_open(fh)};
    
    \draw[-{Latex[length=2mm, width=1.5mm]}, thick, draw=black!50] (cfg1) -- (r1_1) node[midway, right, font=\small\sffamily, text=black!60] {replay};
    
    \node[mpi, below=0pt of r1_1] (r1_2) {MPI\_File\_*\_all(fh)};
    \node[mpi, below=0pt of r1_2] (r1_3) {$\vdots$};
    \node[mpi, below=0pt of r1_3] (r1_4) {MPI\_File\_close(fh)};
    
    \node[posix, below=0pt of r1_4] (r1_5) {open(fd10)};
    \node[posix, below=0pt of r1_5] (r1_6) {write(fd10)};
    \node[posix, below=0pt of r1_6] (r1_7) {$\vdots$};
    \node[posix, below=0pt of r1_7] (r1_8) {close(fd10)};
    
    \node[header] (hn) at (C2) {\underline{Rank$_{n-1}$}};
    
    \node[cfg, below=0.1cm of hn] (cfgn) {CFG$_{n-1}$};
    
    \node[posix, below=0.6cm of cfgn] (rn_1) {open(fdn0)};
    
    \draw[-{Latex[length=2mm, width=1.5mm]}, thick, draw=black!50] (cfgn) -- (rn_1) node[midway, right, font=\small\sffamily, text=black!60] {replay};
    
    \node[posix, below=0pt of rn_1] (rn_2) {write(fdn0)};
    \node[posix, below=0pt of rn_2] (rn_3) {$\vdots$};
    \node[posix, below=0pt of rn_3] (rn_4) {close(fdn0)};
    
    \node[mpi, below=0pt of rn_4]   (rn_5) {MPI\_File\_open(fh)};
    \node[mpi, below=0pt of rn_5]   (rn_6) {MPI\_File\_*\_all(fh)};
    \node[mpi, below=0pt of rn_6]   (rn_7) {$\vdots$};
    \node[mpi, below=0pt of rn_7]   (rn_8) {MPI\_File\_close(fh)};
    
    \node[posix, below=0pt of rn_8] (rn_9) {open(fdn1)};
    
    \path (cfg1.east) -- (cfgn.west) node[midway, font=\Huge, text=black!60] {$\dots$};
    
    \draw[sync] (r0_5.east) to[out=0, in=180] node[midway, sloped, above, font=\tiny\sffamily, text=red!70!black] {} (r1_1.west);
    \draw[sync] (r1_1.east) to[out=0, in=180] (rn_5.west);
    
    \draw[sync2] (r0_6.east) to[out=0, in=180] (r1_2.west);
    \draw[sync2] (r1_2.east) to[out=0, in=180] (rn_6.west);
    
    \draw[sync3] (r0_8.east) to[out=0, in=180] (r1_4.west);
    \draw[sync3] (r1_4.east) to[out=0, in=180] (rn_8.west);
    
\end{tikzpicture}

%% file: imgs/hints_tk.tex
\begin{tikzpicture}[
    box/.style={
        draw=black!70,          
        thick,                  
        rounded corners=2pt,    
        align=left,             
        inner sep=4pt,      
        fill=black!2,           
        font=\scriptsize\ttfamily 
    },
    arrow/.style={
        -{Latex[length=2.5mm, width=1.5mm]}, 
        thick,
        dashed,
        draw=black!70           
    },
    label/.style={              
        font=\scriptsize\sffamily, 
        text=red!70!black,               
        fill=white,             
        inner sep=1pt           
    }
]

\node[box] (mpi) {
\sffamily\bfseries{MPI-IO Tunables}\\[1ex]
"optimization\_params": \{\\
\ \ "use\_collective": 1,\\
\ \ "cb\_block\_size": 1024, \\ 
\ \ "cb\_nodes": 2, \\
\ \ ...
};

\node[box, below=0.3cm of mpi] (posix) { 
\sffamily\bfseries{POSIX Tunables}\\[1ex]
"optimization\_params": \{\\
\ \ "use\_direct": 0,\\
\ \ "disable\_readahead": 0, \\
\ \ "preallocate": \{...\}, \\
\ \ ...
};

\node[box, right=0.6cm of $(mpi.east)!0.5!(posix.east)$] (fbench) {
\sffamily\bfseries{FBench Configuration}\\[1ex]
"file\_config": \{\\
\ \ "file\_name": "/.../file",\\ 
\ \ "total\_repeats": 10,\\
\ \ "method": "write" \\ 
\ \ \textcolor{red!70!black}{\textbf{"hints"}}: \{\textcolor{black!60}{/*injected*/}\},\\
\ \ "chunks": [...] \\ 
\ \ ... \\ 
\}
};

\draw[arrow] (mpi.east) -- ++(0.2,0) node[label, anchor=south, yshift=1pt] {inject} |- ([yshift=0.4cm]fbench.west);
\draw[arrow] (posix.east) -- ++(0.2,0) node[label, anchor=north, yshift=-1pt] {inject} |- ([yshift=-0.4cm]fbench.west);

\end{tikzpicture}

%% file: chapter/4_evaluation.tex
\section{Evaluation}

\subsection{Experimental Setup}
All experiments were carried out on the Corona system at LLNL~\cite{llnl_corona}. The machine has 121 compute nodes, each with an AMD EPYC 7002 processor (48~cores, 256~GB DRAM). It relies on a Lustre file system, an InfiniBand HDR network, and the Slurm/Flux scheduler. Lustre uses progressive striping by default, meaning that files smaller than 64~GB may be distributed across up to 16~OSTs, while larger files are spread over all available OSTs. The stripe size is 1~MB by default, and the system operates in RAID0 mode. Table~\ref{tab:software-v} summarizes the used software.

\begin{table}[tb]
\caption{Software used for the evaluation.}
\label{tab:software-v}
\centering
\small
\begin{tabular*}{\columnwidth}{@{\extracolsep{\fill}} l l l l}
\hline
\textbf{Tool} & \textbf{Version} & \textbf{Tool} & \textbf{Version} \\
\hline
Recorder  & 3.0.0              & IOR      & 4.1.0+dev \\
HACC-IO   & 1.0                & FLASH    & 1.0 \\
Lustre    & 2.15.7\_2.llnl     & UnifyFS  & 2.0 \\
MVAPICH2  & 2.3.7              & LAMMPS   & Stable\_2Aug2023 \\
\hline
\end{tabular*}

\end{table}

To evaluate FBench with the most common I/O patterns of traditional HPC workloads, we selected IOR, HACC-IO, FLASH, and LAMMPS as reference applications. IOR generates synthetic workloads with well-controlled access patterns, HACC-IO represents more realistic application-level I/O behavior, and both FLASH and LAMMPS serve as full production applications.

By design, FBench preserves cache effects in its default configuration, since the resulting measurements reflect the performance users encounter in practice. Caching is therefore enabled throughout the evaluation. When isolation from the cache hierarchy is desired, however, these effects can be eliminated through \texttt{O\_DIRECT}, explicit cache eviction via \texttt{posix\_fadvise(POSIX\_FADV\_DONTNEED)}, and readback reordering that separates the write and read phases. Write fidelity is ensured by issuing \texttt{fsync}/\texttt{MPI\_File\_sync} on close, with every replay executed on freshly created files.


\subsection{IOR (Validation and Scalability)}

\begin{figure*}[tbp]
\centering
\begin{minipage}{0.495\linewidth}
    \centering
    \includegraphics[width=\linewidth]{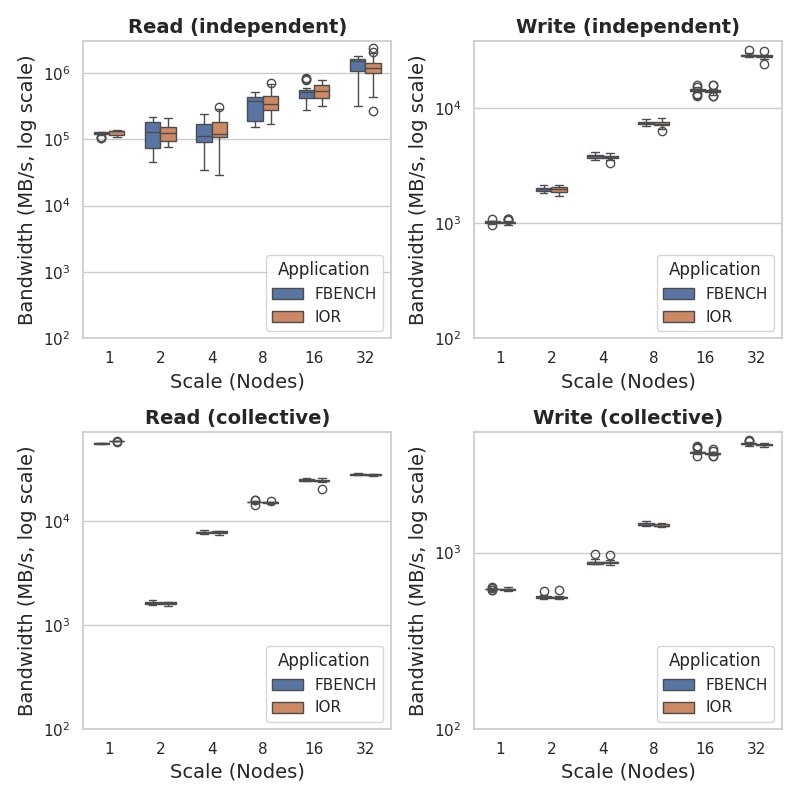}
    \vspace{-15pt}
    \subcaption{MPI-IO Interface}
\end{minipage}
\hfill
\begin{minipage}{0.495\linewidth}
    \centering
    \includegraphics[width=\linewidth]{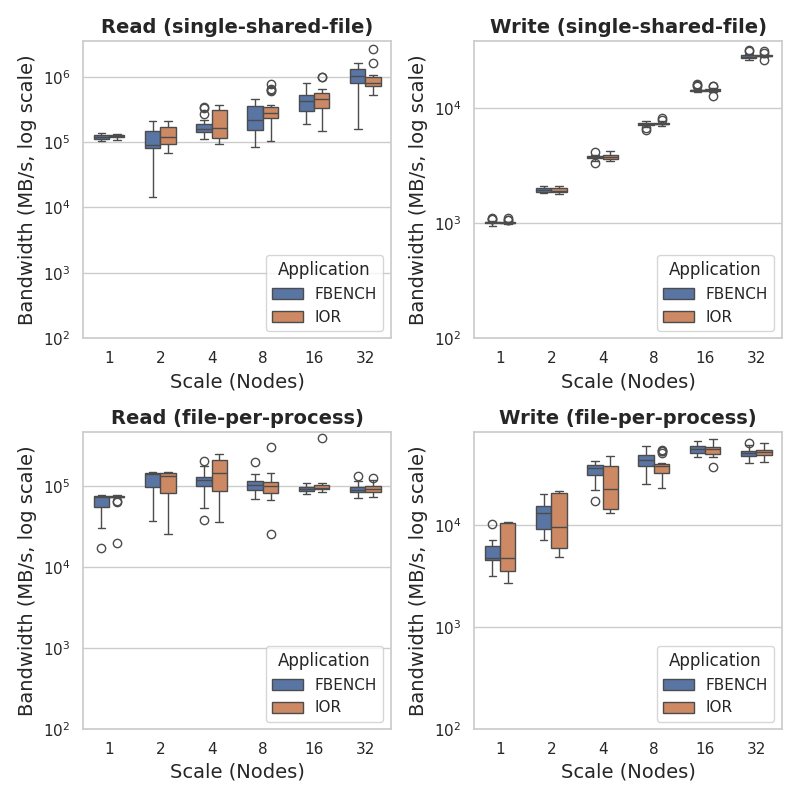}
    \vspace{-15pt}
    \subcaption{POSIX Interface}
\end{minipage}
\caption{Scaling comparison of FBench and IOR up to 1,024 tasks on 32 nodes: As can be seen, the bandwidth increases steadily with the number of nodes across all access patterns, and both tools follow nearly identical trends.}
\label{fig:scaling}
\vspace{-10pt}
\end{figure*}

To assess the accuracy of FBench in reproducing IOR’s I/O characteristics from CFGs, we first instrumented and executed IOR with Recorder. The extracted CFGs were then used as input to FBench to generate the global view, i.e., to derive configuration files and drive the benchmark execution.

We ran IOR with both POSIX and MPI-IO across 1, 2, 4, 8, 16, and 32 nodes, using 32 processes per node. Each process used a fixed block size of 256\,MB and a transfer size of 2\,MB, resulting at the largest scale in a total file size of roughly 264\,GB. The benchmarks covered sequential read and write patterns in two modes for the POSIX interface: File-Per-Process (FPP) and Single-Shared-File (SSF). For MPI-IO, we evaluated both collective and non-collective access for shared files. For each configuration, two runs with 10 iterations each were performed, yielding 20 samples per configuration. The goal is to compare the scaling behavior and bandwidth differences between IOR and FBench under identical settings.

As shown in Figure~\ref{fig:scaling}, FBench and IOR exhibit very similar scaling trends. For MPI-IO independent access, read performance increases steadily and exceeds 2,000\,GB/s at 32 nodes, which is dominated by caching effects, with a deviation of 6.86\%. Write bandwidth follows the same trend, rising to around 30\,GB/s with a deviation of 2.4\%. Collective I/O is even more stable: read operations deviate by only 1.23\%, reaching about 25\,GB/s at the largest scale, while collective writes peak at roughly 4.2\,GB/s with a deviation of 1.51\%.

For the POSIX SSF pattern, read bandwidth spans several orders of magnitude and reaches multi-thousand GB/s at 32 nodes, again driven largely by caching, with a deviation of 6.5\%. Write bandwidth climbs to around 30\,GB/s and differs by less than 1\%. In the FPP pattern, variance is higher but the overall trend remains clear: both tools deliver increasing peak values with growing node counts. Reads surpass 300\,GB/s at the largest scale with a deviation of 1.46\%, while writes reach about 60\,GB/s and deviate by less than 1\%. Overall, FBench reproduces IOR’s scaling characteristics closely. While absolute peak values may diverge slightly, both tools scale almost identically across interfaces and access patterns, even in caching-dominated regions. \looseness=-1

\subsection{HACC-IO (Pattern Validation)}

 \begin{figure}[tb]
    \centering
    \includegraphics[width=\linewidth]{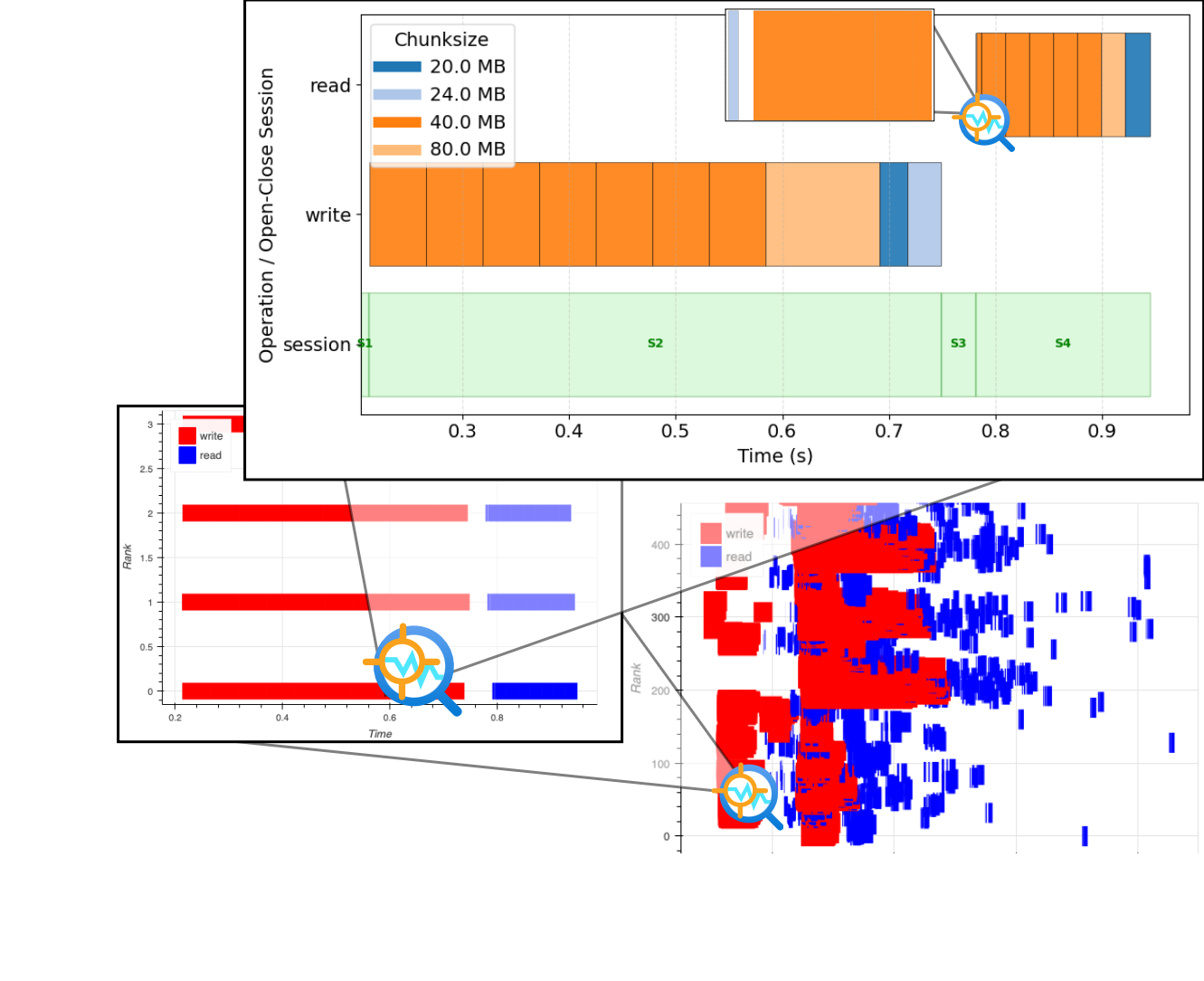}
    \vspace{-40pt}
    \caption{I/O access pattern of HACC-IO on 480 ranks (10 nodes), with more than one million particles per rank.}
    \label{fig:hacc_detailed}
    \vspace{-10pt}
\end{figure}


To demonstrate FBench’s ability to reproduce realistic application I/O patterns, beyond the fixed-size operations of IOR, we used the HACC-IO benchmark. Figure~\ref{fig:hacc_detailed} shows HACC-IO's I/O pattern on 480 ranks, each handling one million particles. It uses a file-per-process pattern: each rank writes its own checkpoint file and later reads it back during the restart phase, resulting in 480 files per cycle.
Examining the sequence for a single rank reveals that the first open-close session performs no actual data I/O. Data is written only in subsequent sessions using blocks of 20, 24, 40, and 80\,MB. The read phase uses the same set of chunk sizes, but the order differs from the write sequence. Reads are shorter and more fragmented, consisting of many small operations placed closely together. In the first read open–close session, only the 24\,MB chunk is accessed, while the remaining chunks are read in a subsequent session. \looseness=-1

\begin{figure}[tbp]
    \centering
    \includegraphics[width=\linewidth]{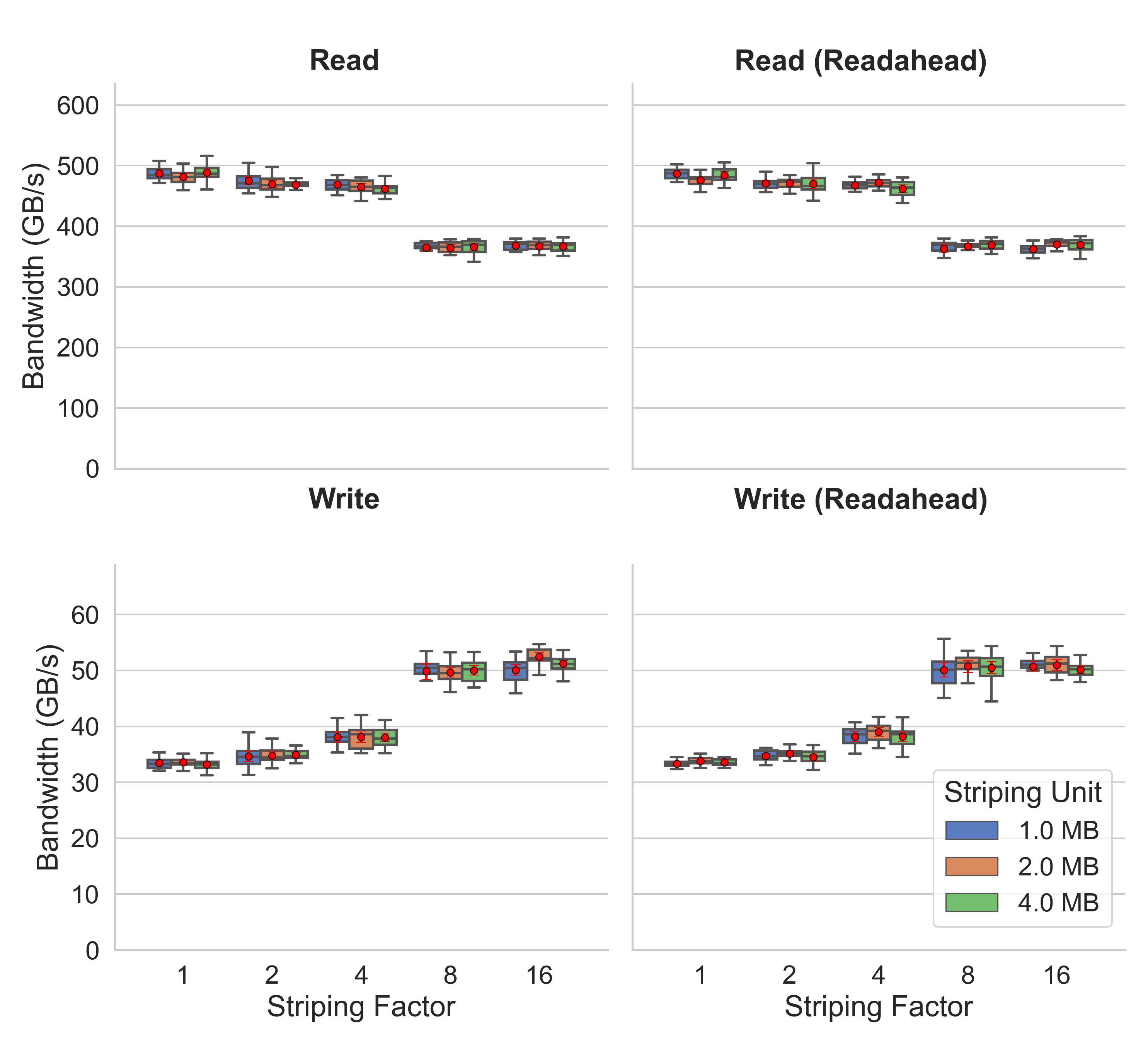}
   \caption{Bandwidth distribution (FBench) by striping factor, striping unit, and enabled readahead for read and write checkpoint files. Mean and $\pm$95\% confidence intervals are highlighted in red. Read performance slightly decreases, while write performance increases with higher striping factors.}
    \label{fig:hacc_fbench}
    \vspace{-15pt}
\end{figure}

This irregular pattern cannot be expressed easily with IOR, which motivates dedicated I/O kernels. In contrast, FBench can reproduce such patterns even via the simplified configuration file. For our experiments, HACC-IO was executed through the POSIX interface and instrumented with Recorder on 10 compute nodes, each running 48 processes and handling one million particles per process. FBench was then run at the same scale, using the trace files to replicate the observed I/O behavior.
Since our analysis focuses on the POSIX interface, we performed a targeted what-if study to examine how specific tunable parameters affect I/O bandwidth. We varied Lustre striping parameters, striping factor up to 16 and striping unit up to 4~MB, which determine how data is distributed across storage targets and influence throughput and contention. Because HACC-IO immediately reads the checkpoint file after writing it, we also varied POSIX-level controls, specifically whether readahead is enabled, which allows the written data to remain cached in the page cache. \looseness=-1
Figure~\ref{fig:hacc_fbench} shows that average read speeds range from approximately 350 to 510\,GB/s, and the distributions are stable across all striping factors. This suggests that performance is dominated by memory and page-cache effects rather than by the underlying disks, explaining why additional I/O-level optimizations provide little benefit. There is no clear winner among stripe units, and enabled versus disabled readahead yields almost identical distributions. With striping factors of 2 and 4, all stripe units show a slight decrease in read bandwidth. For higher striping factors such as 8 and 16, the bandwidth drops significantly. This points to the dip being tied to the stripe count rather than the stripe size. Because paired box plots overlap significantly for all configurations, readahead does not visibly separate the distributions.
In contrast, write performance exhibits a clearer scaling trend. Bandwidth increases with the striping factor from roughly 33-36\,GB/s at factor~1 to over 50\,GB/s at factors~8 and~16. Distributions at higher striping factors also become more consistent, with tighter boxes and fewer extreme outliers. Differences between stripe units remain minor. \looseness=-1

\begin{figure}[tbp]
    \centering
    \includegraphics[width=\linewidth]{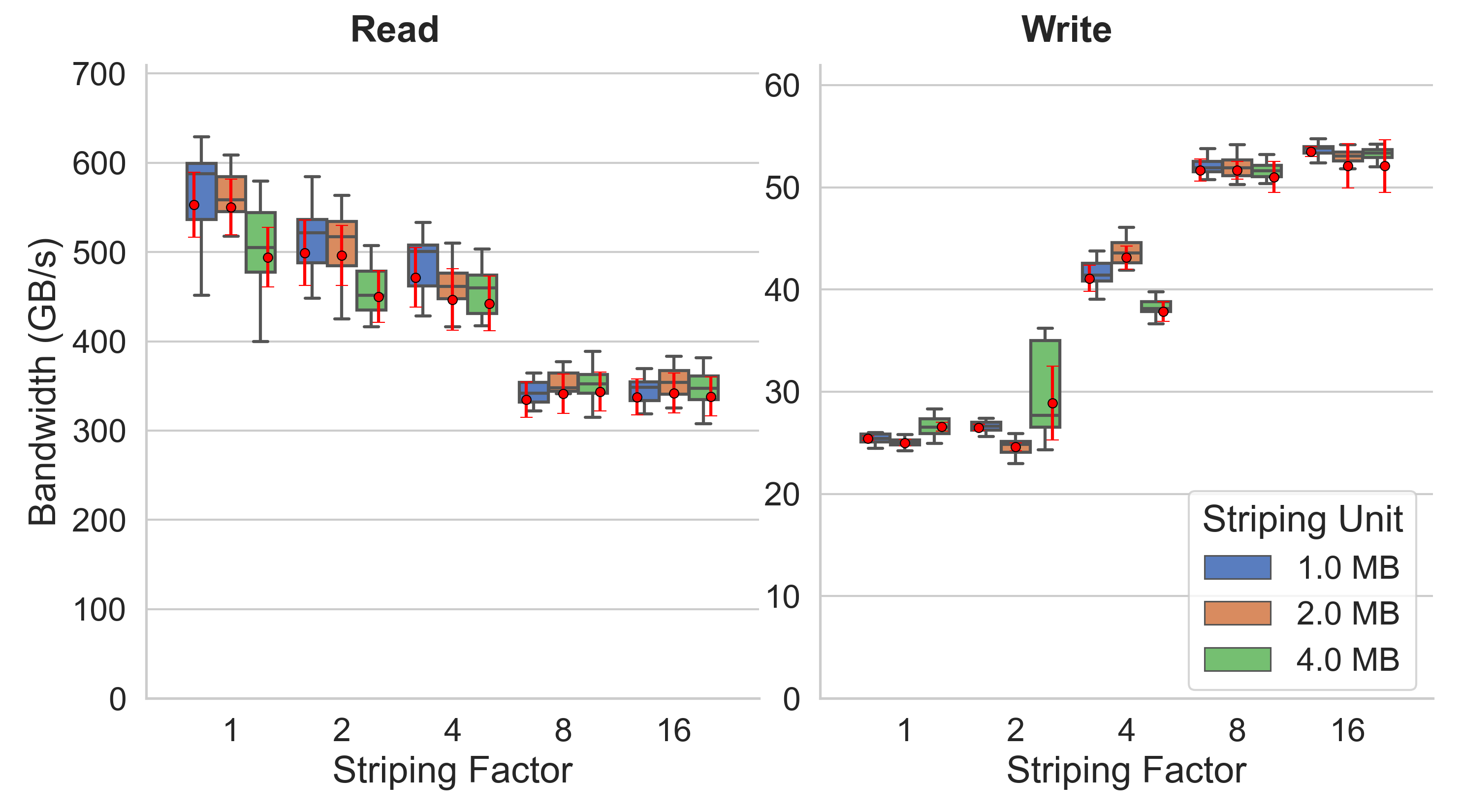}
  \caption{Bandwidth distribution (HACC-IO) by striping factor and striping unit for read and write checkpoint files. Mean and $\pm$95\% confidence interval are highlighted in red. Read performance decreases while write performance increases with higher striping factors. This trend is consistent with Fig. \ref{fig:hacc_fbench}.}
    \label{fig:hacc_origin}
    \vspace{-10pt}
\end{figure}

Because FBench results showed that readahead has little impact and HACC-IO cannot toggle it without code changes, we disabled readahead tuning for the HACC-IO runs. Figure~\ref{fig:hacc_origin} shows the HACC-IO bandwidths when applying the selected Lustre striping configurations. The results match the synthetic what-if analysis, with some deviations in scale and variability. HACC-IO reaches slightly higher peak read bandwidths, but the dip already reappears starting at a striping factor of 2 and becomes very pronounced at 8 and 16. All stripe units follow the same pattern with strong overlap and no clear winner, indicating once again that stripe count, not stripe size, dominates.
Write bandwidth also scales with the striping factor in the original HACC-IO runs. Throughput increases toward factors~8 and~16, while differences between the striping units remain small. Thus, write performance is also primarily driven by the stripe count, with the unit size playing only a secondary role. \looseness=-1

In summary, the what-if analysis with FBench closely reflects the behavior of the original HACC-IO benchmark. Read performance shows the same insensitivity to stripe unit and the same dips at intermediate striping factors, while write performance scales in the same way with the stripe count. Although the original benchmark exhibits slightly less variability, the trends and bottlenecks align well.

\subsection{FLASH}
To evaluate FBench on a real-world large-scale application and conduct a meaningful what-if analysis, we used FLASH Sedov as the target workload. FLASH~\cite{flashx} is a widely used astrophysics and high-energy-density simulation framework designed for large parallel systems. It offers adaptive mesh refinement, modular physics components, and a well-defined I/O mechanism~\cite{Flash-X-SoftwareX}. Depending on the configuration and selected physics modules, FLASH generates a regular, periodic I/O pattern consisting of checkpoint files that capture the full simulation state and smaller plot files for analysis and visualization. Instead of using a reduced I/O kernel, we use the Sedov setup, which models a strong spherical blast wave in a uniform medium~\cite{sedov2018similarity}. This scenario is part of the standard FLASH distribution and produces synchronized write bursts, making it a suitable workload for storage-system tuning studies.
We ran the Sedov experiments on 20 nodes with 40~cores each. Since later evaluations involve applying UnifyFS~\cite{brim2023unifyfs}, which also require resources, we kept the same node layout for all FLASH runs. Each execution produced 21 checkpoint files (each 1193\,MB) and 9 plot files (each 100\,MB), all written collectively as is standard for FLASH. Because FLASH was compiled with the parallel HDF5 interface, which relies on MPI-IO, we pre-loaded the application with Recorder to record MPI-IO calls. From the collected traces, we extracted the global I/O pattern and used the resulting configuration file to drive FBench. Since the pattern is identical for each file type, we limited the what-if analysis to one checkpoint file and one plot file, each executed for 20 iterations with collective I/O enabled. \looseness=-1

 \begin{figure}[t]
    \centering
    \includegraphics[width=0.95\linewidth]{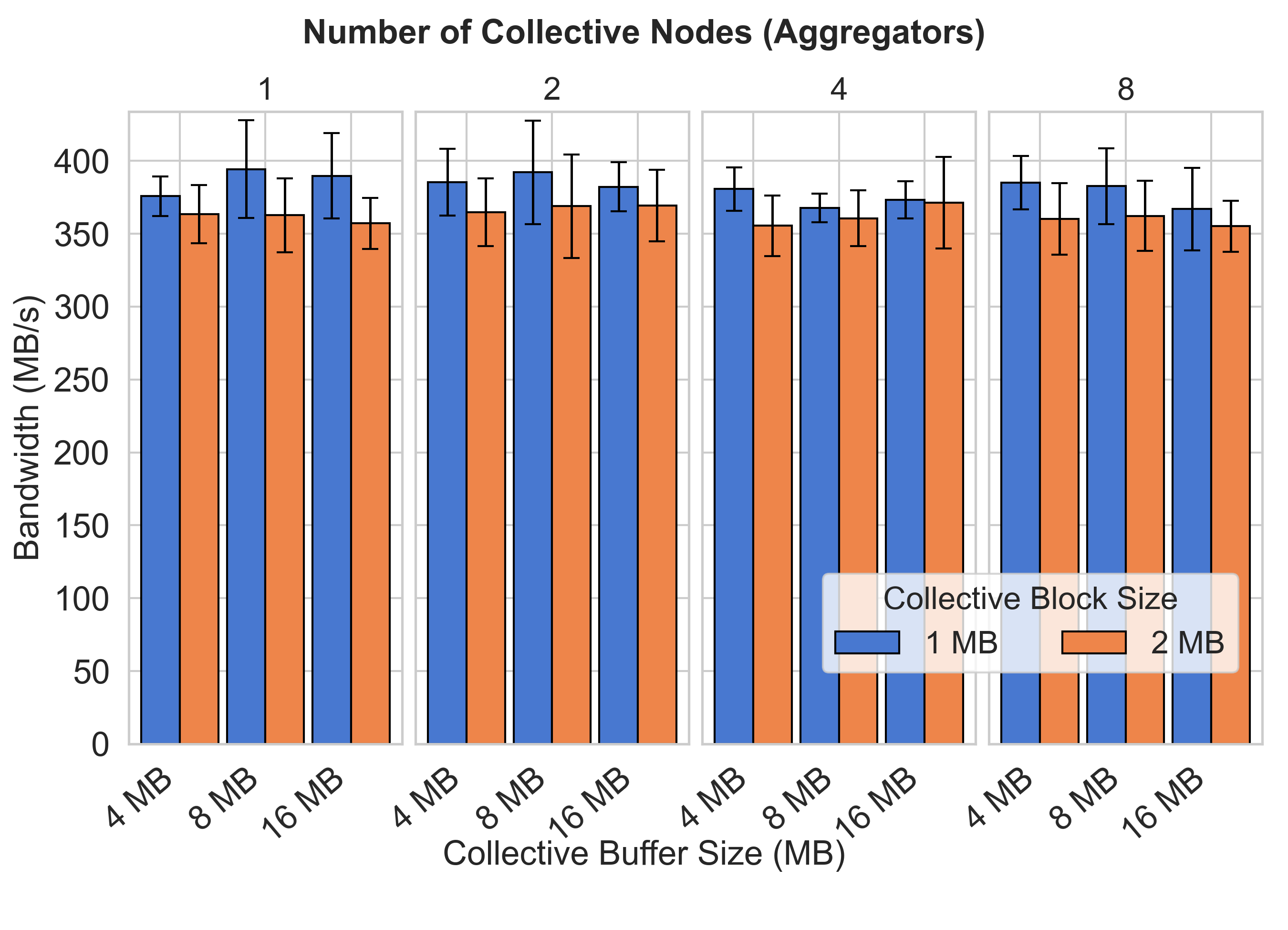}
    \vspace{-10pt}
    \caption{Mean bandwidth (FBench) when writing checkpoint files for various collective buffer sizes, block sizes, and number of nodes with $\pm$95\% confidence interval. Collective buffering shows only a minor impact on bandwidth.}
    \label{fig:flash_cbblock}
    \vspace{-10pt}
\end{figure}

We first evaluated different collective buffering parameters via MPI-IO hints injected into FBench at runtime.
We varied the collective buffer size, the collective buffering block size, and the number of aggregators while keeping Lustre striping fixed (stripe size 0.5\,MB, stripe count 4). Figure~\ref{fig:flash_cbblock} shows the resulting write bandwidths for buffer sizes of 4, 8, and 16\,MB, block sizes of 1 and 2\,MB, and 1, 2, 4, or 8 aggregators. Across all configurations, bandwidth remains within a narrow range of roughly 340-400\,MB/s, indicating limited tuning headroom from MPI-IO collective buffering for this workload. Varying the number of aggregators has minimal effect, and neither increasing the block size nor enlarging the buffer size yields systematic improvements. In fact, larger buffers (16\,MB) tend to produce slightly lower and more variable performance than 4 or 8\,MB, suggesting that the underlying access pattern does not benefit significantly from additional aggregation. \looseness=-1

 \begin{figure}[t]
    \centering
    \includegraphics[width=\linewidth]{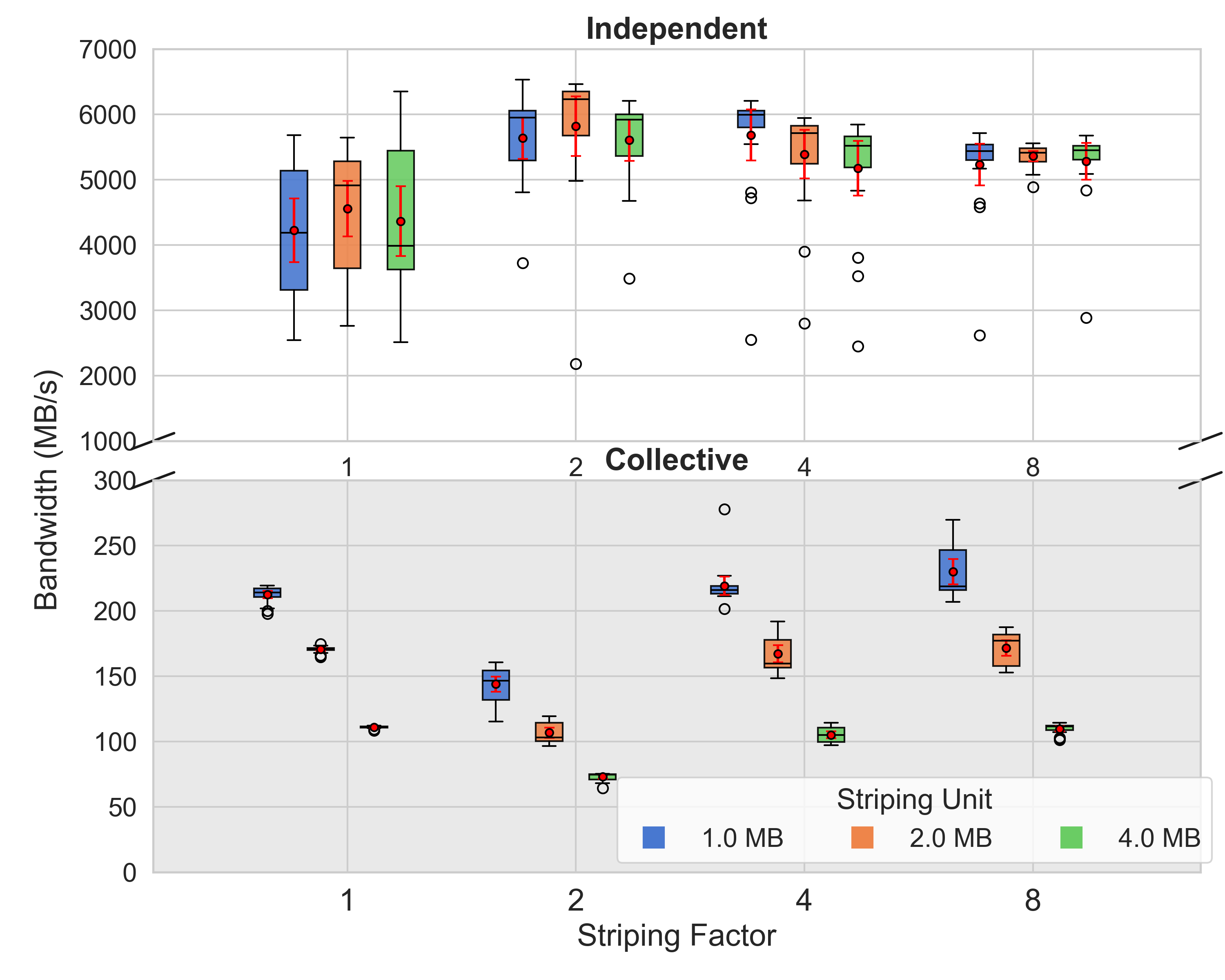}
    \vspace{-15pt}
    \caption{Bandwidth distribution (FBench) by striping factor for write checkpoint files (mean and $\pm$95\% confidence interval in red). Independent I/O consistently achieves higher bandwidth than collective I/O (gray background).}
    \label{fig:flash_lustre}
    \vspace{-10pt}
\end{figure}

In the next step, we studied the effect of Lustre striping settings. Figure~\ref{fig:flash_lustre} shows a pronounced performance gap between collective and independent I/O. Collective I/O yields consistently lower bandwidth across all configurations, independent of the stripe unit or striping factor. In the default configuration with collective I/O enabled, performance drops sharply, and many setups cluster below 200\,MB/s on the log-scaled axis. Increasing the stripe unit further degrades collective I/O, indicating that the collective access pattern interacts poorly with wider stripes on Lustre.

Averaged results confirm this trend: collective I/O achieves only about 180\,MB/s on average, while independent I/O reaches roughly 5.2\,GB/s. Switching from collective to independent I/O thus improves performance by a factor of around $30\times$, showing that the access mode, rather than striping, dominates overall behavior. During the what-if analysis, we also tested disabling Lustre range locking in FBench, but this unexpectedly degraded performance. With newer Lustre version, range locking can only be disabled in combination with direct I/O, which significantly reduces bandwidth. Since FBench uses the same mechanism for disabling Lustre locking as IOR, IOR would encounter the same issue.

\begin{figure}[t]
\centering
\vspace{-5pt}
\begin{minipage}{0.49\linewidth}
    \centering
    \includegraphics[width=\linewidth]{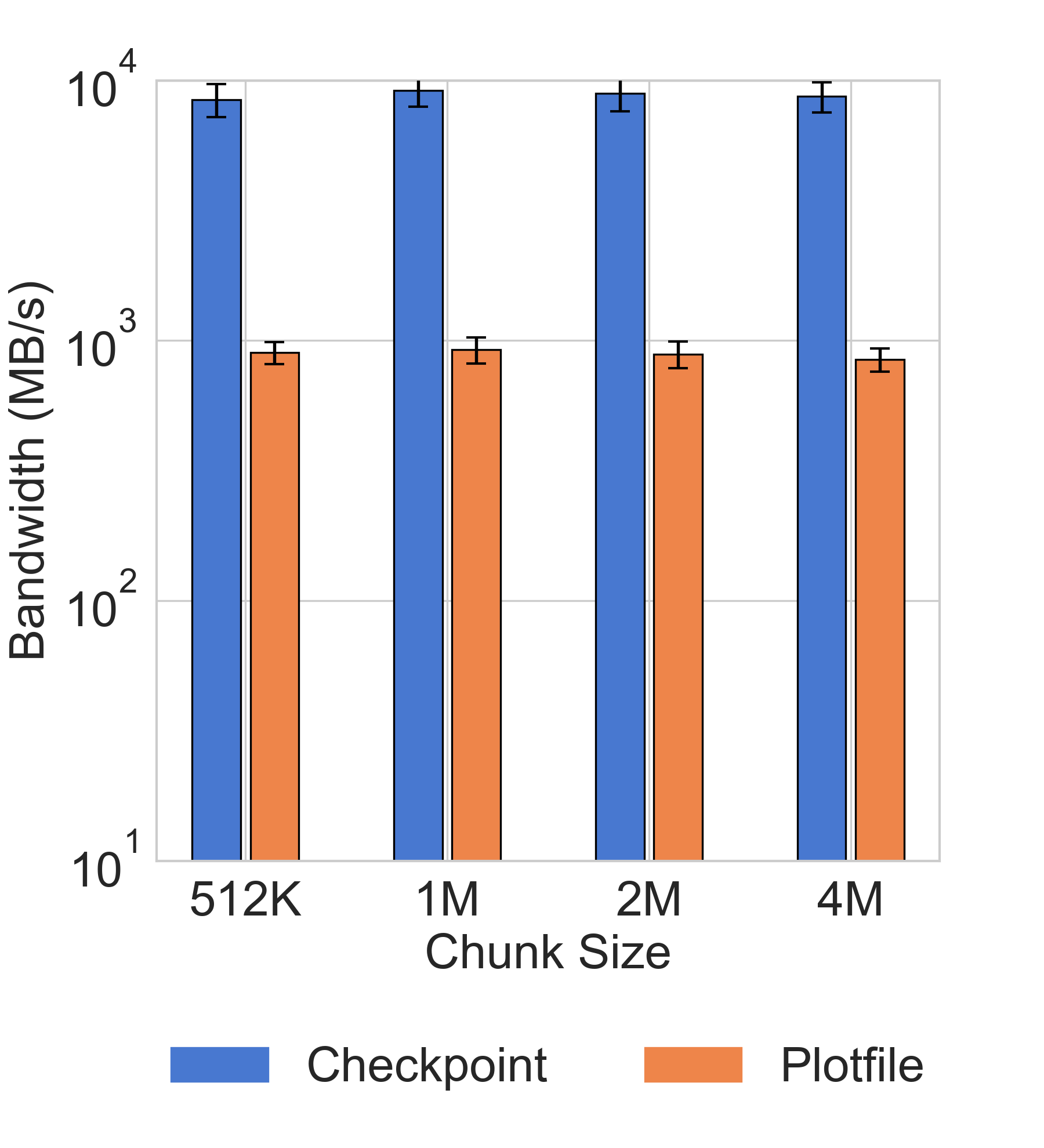}
    \vspace{-20pt}
    \subcaption{FBench}
\end{minipage}
\hfill
\begin{minipage}{0.49\linewidth}
    \centering
    \includegraphics[width=\linewidth]{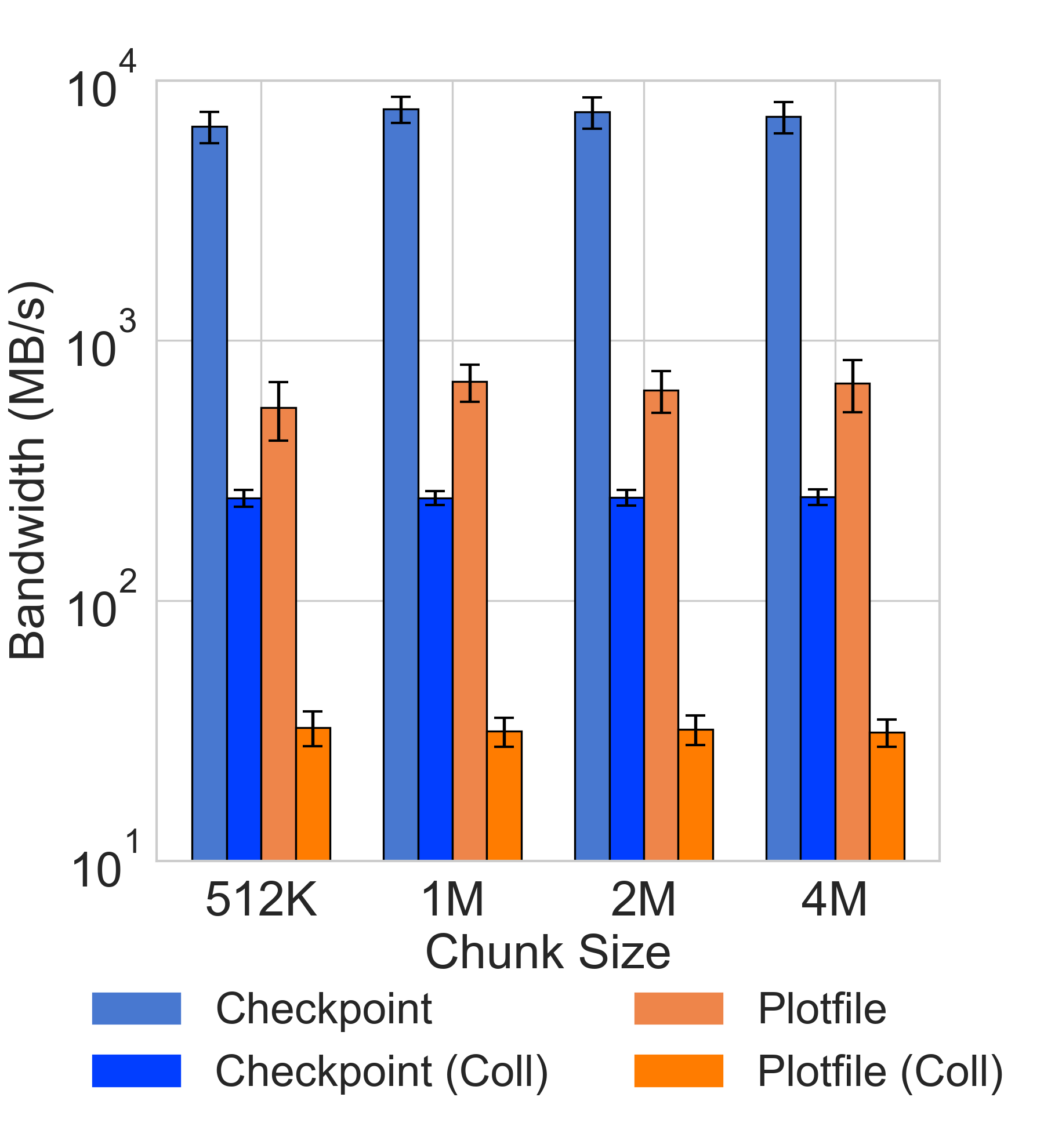}
    \vspace{-20pt}
    \subcaption{FLASH}
\end{minipage}
\caption{Comparison of write bandwidth on UnifyFS: FBench and FLASH exhibit similar performance for independent I/O. FLASH confirms the observation that collective I/O performs significantly worse than independent I/O.}
\label{fig:final_flash}
\vspace{-10pt}
\end{figure}

Finally, we applied FBench to analyze how UnifyFS handles the FLASH workload under different chunk-size configurations (Figure~\ref{fig:final_flash}(a)). A clear performance gap between checkpoint and plot files emerges. Checkpoint write bandwidth remains high across all chunk sizes, typically between 7 and 10\,GB/s with moderate variation. The only notable deviation occurs at a 512\,kB chunk size, which shows a slight reduction. Plot-file performance is substantially lower (around 0.7-1.0\,GB/s) with a much narrower spread, primarily due to the smaller file sizes. Compared to Lustre, the improvement is evident: without any additional tuning, moving from Lustre to UnifyFS increases average independent-write bandwidth from roughly 5.2\,GB/s to about 8~GB/s, an improvement factor of approximately $1.5\times$.
Figure \ref{fig:final_flash}(b) shows bandwidth derived from the original FLASH runs, which aligns well with the FBench measurements. For independent writes, checkpoint bandwidth remains around 8-10\,GB/s across all UnifyFS chunk sizes, while plot-file bandwidth stays in the 0.7-1.0\,GB/s range. This indicates that, despite using a simplified configuration-based pattern, FBench accurately reflects FLASH behavior on UnifyFS for both file types. The results also reinforce the earlier observation that collective I/O consistently underperforms: both checkpoint and plot-file bandwidth drop significantly when collective writes are used, mirroring the trend observed on Lustre in Figure~\ref{fig:flash_lustre}.

\subsection{LAMMPS}
LAMMPS~\cite{plimpton2007lammps} is a classical molecular dynamics code designed for high-efficiency execution on parallel computers. While the \textit{melt} simulation of LAMMPS typically demonstrates the transition of a crystal lattice to a liquid phase, in this work it is used to demonstrate that FBench can significantly reduce time during what-if analysis. 
For this, LAMMPS is running on 10 nodes with 40 processes each, the simulation handles 864 million atoms over 300 timesteps. Via the MPI-IO interface, the simulation performs four collective writes to a shared dump file, with each process contributing approximately 75\,MB (subject to slight per-process variation) and yielding a total dump size of 124\,GB. A deferred flush issued when the file is closed adds a fifth write phase.  Since we want to mimic the I/O behavior of LAMMPS, the original simulation is executed once with the Recorder to obtain traces and thereby generate the configuration file. To obtain the runtime, both LAMMPS and FBench are executed 10 times each. 

The results reveal a substantial difference in execution time for a simulation with only 300 timesteps. While LAMMPS requires on average 566.5\,s, FBench reproduces the I/O workload in 101.8\,s on average. For both measurements, the variance is low. This difference is expected because the runtime of a full scientific application such as LAMMPS depends on multiple factors, including computation, inter-process communication, and other runtime overheads in addition to I/O. FBench, in contrast, isolates and reproduces only the I/O behavior of the application. By eliminating the computational and communication components, FBench requires only about 18\% of the original runtime to reproduce the same I/O workload, therefore significantly accelerating I/O-focused studies and enabling faster what-if analyses. \looseness=-1

Since the I/O patterns of LAMMPS dumps and FLASH's checkpoints are similar, i.e., many processes writing to a single file, we apply also UnifyFS as an optimization. The key difference is that LAMMPS writes to the same file at each interval, whereas FLASH typically writes separate checkpoint files. For the what-if analysis, we evaluate the parameters \texttt{chunk\_size}, which defines the size of the data blocks used for log-based writes, and \texttt{shmem\_size}, which specifies the maximum amount of data buffered in shared memory before being flushed to the underlying storage system. \looseness=-1

Table \ref{tab:lammps:bandwidth_runtime} shows for a shared memory size of 256\,MB a clear difference between the two chunk sizes. While 1 MB chunks achieve a write throughput of about 2.9\,GiB/s, the performance with 4\,MB chunks drops to around 2.3\,GiB/s on average. With 512\,MB of shared memory, both chunk sizes reach the highest measured performance of about 2.96\,GiB/s. At 1 GB of shared memory, performance decreases slightly to around 2.92\,GiB/s, but remains stable. Thus, the results indicate that the chunk size should not be chosen too large when the available shared memory is small. \looseness=-1

\begin{table}[t]
\centering
\caption{Avg. bandwidth and runtime per configuration}
\label{tab:lammps:bandwidth_runtime}
\begin{tabular}{ccc}
\hline

\textbf{Chunk Size} & \textbf{Shared Memory Size} & \textbf{Avg. Bandwidth}  \\
\hline

\rowcolor{black!5}
1 M & 256 M & 2886 MiB/s\\
\rowcolor{black!5}
1 M & 512 M & 2967 MiB/s \\
\rowcolor{black!5}
1 M & 1 G   & 2924 MiB/s \\

\rowcolor{black!10}
4 M & 256 M & 2315 MiB/s\\
\rowcolor{black!10}
4 M & 512 M & 2967 MiB/s\\
\rowcolor{black!10}
4 M & 1 G   & 2913 MiB/s\\
\hline
\end{tabular}
\vspace{-10pt}
\end{table}

As LAMMPS does not provide direct I/O metrics, its performance is assessed using the wall time summary reported for the output step, reflecting the time spent writing dump files. The results show a clear gap between tuned and baseline performance. With UnifyFS, the average output time is around 8.4–8.7\,s, while the Lustre baseline ranges from 64.7\,s to 67.9\,s. Part of this performance gap stems from collective I/O behavior: by default, the aggregator count is coupled to the file system's striping count. In combination with the specific write patterns of LAMMPS, this limits aggregation to only 8 processes even when more nodes are available. Consequently, a mechanism intended to improve I/O efficiency can instead become a scalability bottleneck. Doubling the aggregator count alone reduces the average output time by roughly 20\,s, and with more comprehensive tuning via UnifyFS, an overall output time improvement of roughly $8\times$ is achievable.

In addition to the bandwidth measurements above, the optional inter-I/O-delay mechanism is also evaluated using LAMMPS. These delays are disabled for bandwidth measurements, so that sustained throughput excludes replayed idle time, and enabled whenever faithful temporal reproduction is the objective. The evaluation spans three weak-scaling points (80, 160, and 400 ranks, corresponding to 2N, 4N, and 10N). The recorded trace exhibits five periodic I/O bursts (four data dumps and a single flush at close) overlaid on a scale-invariant compute cadence of roughly 137-139\,s, while the I/O-active fraction of the timeline grows from 13.3\% to 19.6\% as the rank count increases.  With delays enabled, on-the-fly replay preserves the original operation order and reconstructs each idle interval by sleeping for the recorded gap between the completion of one operation and the start of the next. This reproduces all five bursts and matches the inter-burst gaps to within 1-4\% at every scale ($+2.6/+1.1\%$ at 2N, $+1.8/+1.2\%$ at 4N, and $+4.3/+1.8\%$ at 10N). The complete open-to-close timeline deviates by only $-2\%$ at 2N, rising to $+18\%$ at 10N. Because the recorded gaps constitute 80-87\% of this timeline and are reproduced accurately, the residual error is attributable chiefly to the application's deferred flush, which is serialized at close.

%% file: chapter/5_conclusion.tex
\section{Conclusion}
We presented FBench, a flexible benchmark for systematic what-if analysis and I/O performance exploration in HPC. By leveraging CFGs derived from Recorder traces, FBench can either generate simplified global configuration files or replay I/O patterns on-the-fly, supports both POSIX and MPI-IO interfaces, and allows optimization hints to be injected via a JSON configuration, enabling fast experimentation without modifying or rerunning the original application.
Our evaluation shows that FBench accurately reproduces I/O behavior across synthetic and real workloads. For IOR, it closely matches scaling trends across node counts, interfaces, and access patterns. For HACC-IO, it captures the irregular file-per-process pattern and reproduces performance trends under varying Lustre striping settings. For FLASH Sedov, it reveals that collective I/O on Lustre achieves up to $30\times$ lower write bandwidth than independent I/O, largely independent of striping, and that switching from Lustre to UnifyFS improves non-collective write bandwidth by about $1.5\times$ without additional tuning. For LAMMPS, FBench significantly reduces what-if analysis time and enables simple tuning that can reveal up to an $8\times$ improvement. \looseness=-1


Future work includes extending FBench to support additional higher-level I/O libraries, as Recorder already operates at this level, making the inclusion of HDF5, netCDF, and PnetCDF a natural next step. We also plan to quantitatively evaluate timing fidelity, for example via the Wasserstein distance between inter-operation gap distributions, and to develop mechanisms for scaling I/O patterns across different rank counts to enable cross-scale performance extrapolation. Finally, since this work focuses on classical scientific applications with regular checkpoint-restart behavior, we aim to evaluate FBench against more irregular and non-periodic workloads to assess its generality. Overall, FBench provides a scalable, application-agnostic framework that bridges detailed I/O tracing and actionable performance tuning in HPC environments. \looseness=-1